\documentstyle[11pt,twoside,epsfig]{article}    
\graphicspath{{/home/dice/deroeck/papers/f294/}}

\newcommand{\beq}{\begin{equation}}      
\newcommand{\eeq}{\end{equation}}

\newlength{\dinwidth}  
\newlength{\dinmargin}          
\begin{document}      
\noindent     
%===============================titel page================================== 
\begin{titlepage}     
\begin{flushleft}      
       
\noindent     
{\tt DESY 96-039    \hfill    ISSN 0418-9833} \\    
{\tt March 1996} \\          
% {\tt Draft 2.0 March,1st 1996}\\      
% {\tt Editors: G. Bernardi, A. De Roeck and M. Klein}\\
% {\tt Referees: J. Feltesse, U. Straumann}\\
% {\tt Deadline for Comments  8/3/96 12:00}\\ 
% {\tt }\\      
%{\tt eps0470} \\       
%{\tt February, 12. 1996}     \\ 
\end{flushleft}         
\vspace*{3.cm}          
\begin{center}          
\begin{LARGE}           
{\bf A Measurement and QCD Analysis  of \\ the       
Proton Structure  Function {\boldmath $F_2(x,Q^2)$ } at HERA \\ }
%{\bf  A New Measurement of the \\      
%Proton Structure Function  $F_2(x,Q^2)$ at HERA \\ }         

\vspace*{2.cm}          
  H1 Collaboration \\   
\end{LARGE}    
%================================abstract==============================        
\vspace*{4.cm}          
{\bf Abstract:}         
\begin{quotation}
\noindent       
A new %results on a  
measurement of the proton structure function $F_2(x,Q^2)$ is reported       
for momentum transfers squared $Q^2$ between 1.5~GeV$^2$ and  
5000~GeV$^2$ and for Bjorken $x$ between $3\cdot 10^{-5}$ and $0.32$ 
using data collected by the HERA experiment H1 in 1994.      
The data represent an increase in statistics by a factor of ten  
with respect to the analysis of the 1993 data.
%allowing to considerably improve the   accuracy of the measurement. 
 Substantial
extension of the kinematic range towards low $Q^2$ and $x$ has been achieved 
 using dedicated data samples and events 
with initial state photon radiation. 
The structure function
is found to increase significantly with decreasing $x$,         
even in the lowest accessible $Q^2$ region.
The data are well described by a Next to Leading Order QCD fit and the 
gluon density is extracted.        
     
\end{quotation}         
\vfill  
%submitted to Nucl. Phys. B      
\cleardoublepage        
\end{center}   
\end{titlepage}         
%============================authors========================================== 
%-- H1AUTS  Author list by names
%--Status:  5/01/96
\noindent
S.~Aid$^{14}$,                   %HAM2-PD      8/93        Aid
 V.~Andreev$^{26}$,               %LPI -PD                  Andreev
 B.~Andrieu$^{29}$,               %ECPL-PD                  Andrieu
 R.-D.~Appuhn$^{12}$,             %DESY-LEFT  10/95         Appuhn
 M.~Arpagaus$^{37}$,              %ZUTH-LEFT   4/95         Arpagaus
 A.~Babaev$^{25}$,                %ITEP-PD                  Babaev
 J.~B\"ahr$^{36}$,                %ZEUT-PD                  Baehr
 J.~B\'an$^{18}$,                 %KOSI-PD                  Banj
 Y.~Ban$^{28}$,                   %ORSa-ST                  Bany
 P.~Baranov$^{26}$
% ...

%     H1 Institutes as appearing on publications
\noindent
% ...
$\:^2$ III. Physikalisches Institut der RWTH, Aachen, Germany$^ a$ \\
% ...
$ ^{12}$ DESY, Hamburg, Germany$^a$ \\
% ...
$ ^{14}$ II. Institut f\"ur Experimentalphysik, Universit\"at Hamburg,
                                     Hamburg, Germany$^ a$  \\
% ...
$ ^{18}$ Institute of Experimental Physics, Slovak Academy of
                Sciences, Ko\v{s}ice, Slovak Republic$^ f$\\
% ...
$ ^{25}$ Institute for Theoretical and Experimental Physics,
                                                 Moscow, Russia \\
 $ ^{26}$ Lebedev Physical Institute, Moscow, Russia$^ f$ \\
% ...
$ ^{28}$ LAL, Universit\'{e} de Paris-Sud, IN2P3-CNRS,
                            Orsay, France\\
 $ ^{29}$ LPNHE, Ecole Polytechnique, IN2P3-CNRS,
                             Palaiseau, France \\
% ...
$ ^{36}$ DESY, Institut f\"ur Hochenergiephysik,
                              Zeuthen, Germany$^ a$\\
 $ ^{37}$ Institut f\"ur Teilchenphysik,
          ETH, Z\"urich, Switzerland$^ i$\\
% ...
$ ^a$ Supported by the Bundesministerium f\"ur Bildung, Wissenschaft,
        Forschung und Technologie, FRG, under contract numbers
        6AC17P, 6AC47P, 6DO57I, 6HH17P, 6HH27I,
        6HD17I, 6HD27I, 6KI17P, 6MP17I, and 6WT87P \\
% ...
$ ^f$ Supported by the Deutsche Forschungsgemeinschaft\\
% ...
$ ^i$ Supported by the Swiss National Science Foundation\\

\newpage 

\section{Introduction}  
A prime task of the electron-proton collider HERA is the 
investigation of the structure of the proton. 
Measurements of the inclusive lepton-proton 
scattering cross section have been crucial for
the understanding of  proton  substructure~\cite{eisele}.       
Early electron-proton scattering  experiments have discovered
   pointlike proton constituents by observing a scale invariant dependence
of the proton structure function $F_2(x,Q^2)$ on the four-momentum
  transfer squared 
$Q^2$ at  Bjorken $x \geq 0.1$ and $Q^2$ values of about 5
  GeV$^2$. 
Subsequent neutrino scattering experiments have established
  the Quark Parton Model (QPM)
       as a valid picture of the valence and sea quarks as
   constituents of the proton. The interaction of these partons as
 mediated by gluons is successfully described by Quantum Chromodynamics (QCD)
 which has been tested with high precision in muon-nucleon 
 deep-inelastic scattering (DIS)
experiments.
 Experiments at HERA  extend   the previously 
accessible kinematic range up to very large squared momentum transfers,        
$Q^2 >   10^3$~GeV$^2$, and down to very small values of Bjorken     
$x <   10^{-4}$.       

The first measurements of the structure function $F_2(x,Q^2)$
  reported at HERA, based on data collected in 1992,
 revealed  its strong 
rise  at low $x < 10^{-2}$ with decreasing  
$x$~\cite{H193,ZEUS93}.
This rise was  confirmed with the more precise data of 1993~\cite{H194,ZEUS94},
based on  an order of magnitude increase in statistics.
Such a behaviour is qualitatively expected in the asymptotic
limit of Quantum Chromodynamics~\cite{Alvaro}. It is, however, not    
clear whether the rise of  
$F_2$ is fully described by the linear QCD evolution equations, such as the  
conventional DGLAP evolution~\cite{DGLAP} in $\log Q^2$ 
or by the BFKL         
evolution~\cite{lipatov} in $\log(1/x)$,  or    
whether there is a significant effect due to  
      non-linear parton recombination \cite{GLR}.    
Furthermore, it is also unclear whether this rise will persist at low
values of $Q^2$      of the order of one GeV$^2$. For example,           
Regge inspired models expect $F_2$ to be rather    
flat as function of $x$ at small $Q^2$.        
The     
quantitative investigation of the quark-gluon interaction  dynamics at         
low $x$ is one of the major challenges at HERA. It           
requires  high precision for the 
$F_2$ measurement and 
complementary investigations of the characteristics 
of the hadronic final state~\cite{BFKL}.

In this paper an analysis is presented of inclusive deep-inelastic    
scattering  data taken by the H1 collaboration in 1994 with      
an integrated luminosity of $2.7$~pb$^{-1}$, which is   
 an order of magnitude larger than        
in 1993. 
The incident electron\footnote
{HERA operated with
 $e^-p$ collisions in 1992, 1993 and the start of 1994, and
$e^+p$ collisions for the major part of 1994. In this paper the incident
and scattered lepton will  always be referred to as an ``electron''.}
energy $E_e$ was  27.5 GeV and the     
proton energy $E_p$ was 820 GeV. 
The accessible kinematic         
range has  been extended to the very high $Q^2$          
region and the  structure function $F_2$ has     
been investigated at a new level of precision.
To reach lower $Q^2$ values and correspondingly lower $x$ values,
special  samples were analysed of events with shifted interaction vertex, 
 and of 
 events with tagged initial state photon radiation.

This paper is organized as follows. After a short
introduction to the kinematics of  inclusive $ep$ scattering 
(section 2), the H1 apparatus is briefly sketched (section 3).
The different 1994 data samples, the luminosity determination
 and the Monte Carlo
simulation are described in section 4. 
Next the event selection including the background rejection
(section 5) is discussed for the different data samples 
used.
Section 6 describes the $F_2$ analyses. In section 7 the
results are discussed.
A phenomenological analysis of $F_2$ is performed and the data are compared to recent 
model calculations at low $Q^2$.  The  data are also 
studied in the framework of 
perturbative QCD and the gluon distribution is extracted.
The paper is summarized in section 8.
     
%--------------------------------------------------------------------
\section{Kinematics}
%--------------------------------------------------------------------
The structure function $F_2(x,Q^2)$ is derived from the inclusive     
electron-proton scattering 
cross section. It depends on the squared four-momentum 
transfer $Q^2$ and the scaling variable $x$. These variables 
are related to the inelasticity parameter $y$ and to the total squared         
centre of mass energy of the collision $s$ since $Q^2= xys$ with $s= 4   
E_e E_p$.      
%The data were recorded with the H1   
%detector, which is described in \cite{h1detec}.    
A salient feature of the HERA collider    
experiments is the possibility of measuring not only the scattered    
electron but also the complete hadronic final state, apart from losses         
near the beam pipe.     
This means that the kinematic variables $x,~y$ and $Q^2$   
can be determined with complementary methods which           
 are  sensitive to different systematic effects.        
These methods were exploited and detailed already in \cite{H194}    
which describes the analysis          
of the 1993 data.       
%The comparison  of the results obtained with different methods        
%improves  the accuracy of the $F_2$ measurement. 
An appropriate
combination of the results ensures maximum coverage of the available  
kinematic range.        

The methods used in the analysis of the 1994 data are  the 
so called ``E'' (electron)  method     
 using only the information of the scattered   
electron and           
 the so called ``$\Sigma$'' method        
%$(\Sigma)$
 calculating the kinematics based on both the scattered electron
 and the  hadronic final state measurements~\cite{sigma}.
The E method, which is independent of the hadronic final state, 
apart from the requirement that the interaction vertex is reconstructed
using the final state hadrons, has at large $y$ the best resolution
in $x$ and $Q^2$ but needs sizeable radiative corrections.
At low $y$ the E method is not  applied due to the degradation
of the $y_e$ resolution as $ 1/y$.    
The $\Sigma$ method, which has small radiative corrections, relies
mostly on the hadronic measurement which has still an acceptable
resolution at low $y$ values and can be used from very low
to large $y$ values.
%The E method has the best resolutions in $x$ and
%$Q^2$ at large $y$ and is independent of the hadronic final state,   
%apart from the requirement that the 
%interaction vertex is reconstructed using the final state hadrons. 
%At low $y$ the  E method  is not  applied due to degrading resolution
%($\sim 1/y$).
%The $\Sigma$ method relies mostly on
%the hadronic measurement which has still an acceptable resolution at
%low $y$ values and can therefore be used from very low to large   
%$y$ values.      
The E and $\Sigma$ results were compared in order to control the calculation of
the systematic errors.
The basic formulae for $Q^2$ and $y$ 
 for the E method are:
\begin{equation}
  y_e   =1-\frac{E'_e}{E_e} \sin^{2}\frac {\theta_e} {2}
   \hspace*{2cm}
   Q^2_e = 4E'_eE_e\cos^2\frac{\theta_e}{2} 
= \frac{E^{'2}_e \sin^2{\theta_e}}{ 1-y_e} \label{kinematics1}
\end{equation}
where $E'_e$ and $\theta_e$
are the energy and polar angle of the scattered electron.
The polar angle $\theta_e$ is  defined with respect to     
 the proton beam or $z$ direction, termed ``forward" region.   
The formulae for the $\Sigma$ method
are
%    are constructed requiring  $Q^2$ and $y$
%to be independent of the incident electron energy.
%Using the conservation of the total $E-P_z \equiv
%       \sum_i{E_i-p_{z,i}}$, the sum extending over
%all particles of the event,  $2E_e$ is replaced by
%  $  \Sigma + E'_e(1-\cos{\theta_e})$  which gives
\begin{equation}
   y_{\Sigma} = \frac{\Sigma}{ \Sigma + E'_e(1-\cos{\theta_e})}
   \hspace*{2cm}
   Q^2_{\Sigma} = \frac{E^{'2}_e \sin^2{\theta_e}}{ 1-y_{\Sigma}}
\end{equation}
with
\begin{equation}
   \Sigma=\sum_h{(E_h-p_{z,h})}.
%   \hspace*{2.cm}
% p^{h~2}_T=(\sum_h{p_{x,h}})^2+(\sum_h{p_{y,h}})^2
\end{equation}
Here   $E_h$ and $p_{z,h}$ are the energy and longitudinal momentum component
of a particle $h$, the summation is  over all hadronic final
state particles and the masses are neglected.
The denominator of $y_{\Sigma}$ is equal to $2E_e$ but measured with all
secondary particles.
Thus
\begin{equation}
   y_{\Sigma} = \frac{y_h}{ 1+y_h-y_e} \label{kinematics2}
\end{equation}
with the standard definition
\begin{equation}
   y_h = \frac{\Sigma}{2E_e}.\label{kine4}
\end{equation}
The variable $x$ is calculated as
$x=Q^2/ys$.
%, with $s$ the centre of mass energy squared of the 
%$ep$ collision.
%check      
%that the systematic errors  were correctly evaluated.        
%All these  methods        were  utilized to measure $F_2$.
%The resolutions  of the reconstructed
%$x$ and $Q^2$ variables  are discussed at the end of section 5.
%
%\end{document}       
       
%-------------------------------------------------------------------- 
%--------------------------------------------------------------------
\section{The H1 Detector}
%--------------------------------------------------------------------

The H1 detector~\cite{h1detec} is a nearly hermetic multi-purpose apparatus
built to investigate the inelastic high-energy interactions
    of electrons
and protons at HERA. The structure function measurement relies
essentially on the
inner tracking chamber system and on the backward electromagnetic and
the liquid argon calorimeters which will be described here briefly.
 
The tracking system includes the central tracking chambers,
the forward tracker modules and a backward proportional chamber.
These chambers are placed around the beam pipe at $z$ positions
between --1.5 and 2.5 m. A superconducting solenoid
 surrounding both the tracking system and the liquid argon
calorimeter provides a uniform magnetic field of 1.15~T.
 
The central jet chamber (CJC) consists of two concentric drift chambers
covering a polar angle range from  $15^o$ to $165^o$. Tracks
crossing the CJC are measured with a transverse momentum resolution of
${\delta p_T}/{p_T}< 0.01 \cdot p_T/$GeV.
The CJC is supplemented  by two
cylindrical drift chambers at radii of  18 and 47 cm, respectively,
         to improve the 
determination of  the $z$ coordinate of
the tracks.  A proportional chamber is
attached to each of the $z$ drift chambers for triggering. 
%The inner one (CIP) was used here
%to estimate residual photoproduction background.
 
A tracking chamber system made of three identical modules measures
hadrons emitted in the forward direction ($7^o$ to $20^o$). The forward
tracker (FT) is used to determine  the vertex
for the events which leave no track in the CJC.
This allows an extension of the analysis to  larger $x$ values.
 
In the backward region, attached to the backward electromagnetic
calorimeter (BEMC),
 a four plane multiwire proportional chamber (BPC)
was located with a  polar angle acceptance
of $151^o$ to $174.5^o$. The
BPC provides a space point for charged particles entering the BEMC
which is used for low $Q^2\leq 120$ GeV$^2$
events to identify electrons and to
measure $\theta_e$. The spatial resolution for reconstructed BPC hits
is about 1.5 mm in the  plane perpendicular to the beam axis.
 
The backward  electromagnetic calorimeter~\cite{BEMC} which detects the
scattered electron at low $Q^2$   is made of 88
lead/scintillator stacks with a size of $16 \times 16$ cm$^2$ and a depth
of 22 radiation lengths corresponding to about one interaction length.
Around the beampipe the stacks are of triangular shape.
The  angular coverage of the BEMC is $155^o<\theta_e<176^o$.
A 1.5 cm spatial
        resolution of the lateral shower position is achieved
using four  photodiodes which detect the wavelength shifted light from
each of the scintillator stacks.
A scintillator hodoscope (TOF) situated behind the BEMC is used to veto
proton-induced background events based on their early time of arrival
compared with nominal $ep$ collisions.

Hadronic final state energies and the scattered electron at
high $Q^2$ ($Q^2 \ge 120$ GeV$^2$) 
  are measured in the
liquid argon (LAR) calorimeter~\cite{LARC} which covers an  angular
region between $3^{o}$ and $155^{o}$. The calorimeter consists of an
electromagnetic 
       section with lead absorber plates and a hadronic section with
stainless steel absorber plates. 
Both sections are highly segmented in the transverse and longitudinal
directions with about 44000 cells in total.
The electromagnetic part has a depth between
20 and 30 radiation lengths. The total depth of both calorimeters
varies between 4.5 and 8 interaction lengths. 
% 
%Hadronic final state energies and the scattered electron at
%high $Q^2>120$ GeV$^2$ 
%  are measured in the highly segmented
%liquid argon calorimeter (LAr) \cite{LARC} which covers an  angular
%region between $3^{o}$ and $155^{o}$. The LAr consists of an
%electromagnetic 
%       section with lead absorber and a hadronic section with
%stainless steel absorber. The electromagnetic part has a depth between
%20 and 30 radiation lengths. The total depth of both calorimeters
%varies between 4.5 and 8 interaction lengths. 
%The most backward part of
%the LAr is a small electromagnetic calorimeter (BBE) which covers the
%polar angle range from $150^o$ to $155^o$.

The luminosity was determined from the measured cross section of
      the Bethe Heitler (BH)
reaction $ep \rightarrow ep\gamma$. The final state electron and photon
can be  detected in calorimeters
(electron and photon ``taggers") close to the beam pipe but at  large
distances from the main detector (at $z=-33$ m and $z=-103$ m).

\section{Data and Monte Carlo Samples}

\subsection{Data Samples}

\begin{figure}[htbp]\unitlength 1mm   
\begin{center}          
\epsfig{file=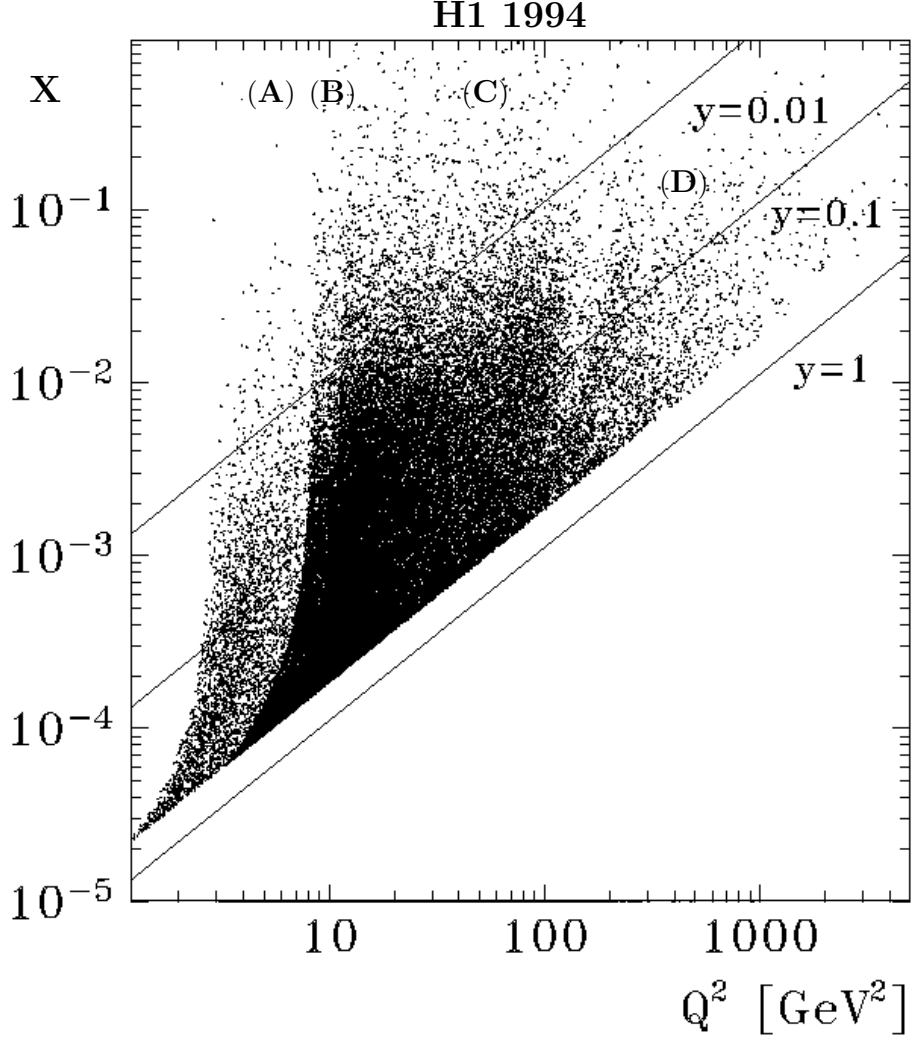,width=16cm,
bbllx=0pt,bblly=450pt,bburx=525pt,bbury=645pt,angle=90.} 
\begin{picture}(0,0) \put(-30,150){{\Large \bf H1 1994}} \end{picture}
\begin{picture}(0,0) \put(-56,140){({\large \bf A})} \end{picture}
\begin{picture}(0,0) \put(-49,140){({\large \bf B})} \end{picture}
\begin{picture}(0,0) \put(-30,140){({\large \bf C})} \end{picture}
\begin{picture}(0,0) \put(-5,128){({\large \bf D})} \end{picture}
\begin{picture}(0,0) \put(-90,140){{\Large \bf X}} \end{picture}
\end{center}   
\caption[]{\label{q2xplot}       
\sl Distribution of the  event sample in the $(x,Q^2)$   
plane. The 4 visible regions (A,B,C,D) correspond to A) events recorded during
a period in which the interaction region was shifted  with respect to  the 
nominal position allowing access to larger $\theta_e$; B) events from
the nominal vertex position taken in a period in which the 
innermost BEMC stacks of triangular shape
were included  in the trigger (``opened
 triangles'', see text) or C) not included; D)
high $Q^2$ events with the scattered electron detected in the 
LAr calorimeter.
}
\end{figure}

Several data samples have been analysed in order to cover 
  maximally  the kinematic plane.
 The distribution of the events is
shown in Fig.~\ref{q2xplot}.  
The majority of the
events  are produced 
 with the interaction vertex centered around zero in $z$,
 called the ``nominal vertex'' sample (shown as regions  C  and D 
%-- high $Q^2$ --
in Fig.~\ref{q2xplot}).
Throughout this paper, the low (high) $Q^2$  sample refers to events in
which the scattered electron has been detected in the BEMC (LAr
calorimeter). 
% The analysis of the low $Q^2$  nominal
%vertex  sample has been based only on data taken under the best
%experimental conditions~\cite{pineiro} in order
%to optimize the precision of the measurement. 
%To reduce the systematic errors of the $F_2$ measurement, a strict
%data selection was performed based on the behaviour of the main
%detector components. 
%This behaviour was required to be optimal for
%the low $Q^2$ analysis of the nominal vertex sample which allows to
%reach the highest precision.
%, bringing the luminosity used
%in this analysis to a value of  2.2~pb$^{-1}.
%        
To reduce the systematic errors of the $F_2$ measurement, a strict
  data selection was performed based on the behaviour of the main
  detector components. This behaviour was required to be optimal for
 the low $Q^2$ analysis of the nominal vertex sample which allows    
 the highest precision to be reached.     
The remaining integrated
luminosity for the low $Q^2$ sample is 2.2~pb$^{-1}$, the one for the 
high $Q^2$ sample is 2.7  pb$^{-1}$.
%To reduce the systematic errors of the $F_2$ measurement,    
%a strict data taking period 
%selection was performed based on the  behaviour         
%of the main detector components. 
The number of accepted events per unit luminosity  
was checked to be constant within         
statistical errors during the data taking period.    

 In order to study the  behaviour of $F_2$ at small $Q^2$ several
 means  were used to extend the acceptance to this kinematic region
 using  special event samples.        
 For DIS events at very low $Q^2$  
the electron is scattered through a large  
  angle $\theta_e$.       
 For $\theta_e$ values greater than $173^0$ 
and the interaction vertex at its nominal position at $z = +3$ cm 
the electron
hits the inner edge   of the BEMC calorimeter or 
remains  undetected near the beam pipe.
   The acceptance extension  in the backward region was realized as follows:  
 \begin{itemize}
 \item
  During  good accelerator background conditions,         
  the innermost   parts
 of the backward electromagnetic calorimeter (BEMC)
 around the beampipe were
 included in the trigger for part of the time. Since these  detector elements are of 
 triangular shape,       
 these data will be referred to as the ``open triangle"   
 data sample. An integrated luminosity of 0.27~pb$^{-1}$ 
 was accumulated.   
%  under these conditions. 
The kinematic region covered by this sample
 is shown as region (B) in Fig.~\ref{q2xplot}.  
 \item As in 1993~\cite{H194},
  the interaction point was shifted 
 in the forward direction to an average  position of $z= +67$ cm    
 which permits measurements up to $\theta_e \simeq 176.5^o$.
 This sample of 58 nb$^{-1}$ of data
 is referred to as the ``shifted vertex'' data sample to distinguish it from 
the data with a nominal event vertex.
It covers region (A) in Fig.~\ref{q2xplot}.         
 \item  
 The low $Q^2$ region was also accessed   by analyzing events       
  from the so called     
 early proton satellite bunch
 colliding with an electron bunch at $z \simeq +68$~cm. The kinematic region covered by this sample is similar to that of the
 shifted vertex data sample. The  ``satellite" data sample          
 amounts to $\simeq 3\%$ of the
 total data corresponding to a    total ``luminosity" of       
 68 nb$^{-1}$  selected over the whole run period.        

% is shown as region (a) in Fig.~\ref{q2xplot}. 
 \item
 Finally, a sample of  deep-inelastic radiative 
 events was extracted
 with a hard photon emitted collinear with the incident electron. 
 These events  have a reduced incident electron beam energy
 which allows access to very low 
   $Q^2$ values with the present detector setup.
% A first experimental study of this process together with other hard photon
% radiative processes has been recently published\cite{h1radpap} by the H1 
% collaboration 
% using its 1993 data.
 Since only about 2\% of the DIS events are tagged as  radiative events, 
 the nominal vertex sample had to be used for this study. 
 Subsequently the tagged radiative events are referred to
 as ``the radiative
 event sample'' and the 
bulk of the data are sometimes called ``non-radiative" in
  contrast.
\end{itemize}

 \subsection{Luminosity Determination}

% An important global systematic effect on the 
% structure function is the 
% precision of the 
% luminosity measurement. 
 The most precise method of determining the luminosity from the
  reaction $ep \rightarrow ep\gamma$
 is based on the measurement of the energy spectrum of hard photons
 ($E_{\gamma}> 10$ GeV) as explained  in~\cite{h1radpap} for the 1993 data.
 The main  uncertainties  of the measurement of the integrated luminosity
for the 
 1994 nominal vertex data   are: 
 the photon tagger absolute energy scale (0.9\%), the trigger efficiency
 of the luminosity system
 (0.3\%), the precision of the electron gas background subtraction
 (0.4\%), the photon-tagger acceptance (0.5\%), multiple photon overlaps
 (0.4\%), the precision of integration resulting from the 10 sec interval between
consecutive luminosity measurements
 (0.5\%) and the correction for satellite
 bunches (0.5\%). Major improvements with respect to 1993 data 
include the trigger 
 efficiency, the satellite bunch correction and the precision of the energy scale
 in the photon tagger.
 The  precision of the 
 luminosity measurement for the nominal vertex  data sample  
 is 1.5\% 
 which represents an improvement of a factor of 3 with respect to
  the 1993 data analysis. For the shifted vertex data 
 sample the luminosity uncertainty       
 is 3.9\%. 

 The results of this measurement  were  checked for consistency
with a sample 
 of Bethe Heitler
 events in which both  the electron and photon are detected 
 simultaneously, and with QED Compton events. Both these analyses are subject 
 to different systematics, compared with the hard photon method, allowing a 
 cross check of the luminosity with a precision of up to 6\%.

  The integrated luminosity         
 of the satellite data sample was obtained from  the measured 
 integrated luminosity      
 for the shifted vertex data multiplied by the efficiency corrected    
 event ratio in a kinematic region common to both data sets.  
 The precision of that luminosity determination was estimated to be  
  7.1\%.   

 \subsection{Monte Carlo Simulation}

 More than one million  Monte Carlo DIS events were generated using           
 the DJANGO~\cite{django} program.  
 The Monte Carlo event statistics
  correspond to an integrated luminosity of                 
 approximately $18$~pb$^{-1}$. The DJANGO 
 program is based on                          
 HERACLES~\cite{heracles} for the electroweak interaction                     
 and on  the LEPTO program~\cite{lepto} 
 to simulate the hadronic             
 final state. 
HERACLES includes first order radiative corrections,            
 the simulation of real Bremsstrahlung photons and the longitudinal           
 structure function. The acceptance corrections were performed using the      
 GRV parametrization~\cite{NEWGRV} 
which describes rather well the HERA         
 $F_2$ results based on the 1993 data. % for $Q^2 > 8.5$~GeV$^2$. 
 LEPTO uses the colour dipole model (CDM) as implemented in                   
 ARIADNE~\cite{cdm} which is in good agreement with data on the energy        
 flow and other characteristics of the final                                  
 state as measured by H1~\cite{h1flow} and ZEUS~\cite{zeflow}. For the        
 determination of systematic errors connected with the topology                  
 of the hadronic final state,     the HERWIG model~\cite{herwig} was          
 used.
% in a dedicated analysis.

 Photoproduction background was simulated based on the PHOJET~\cite{phojet},
 PYTHIA~\cite{pythia} and RAYVDM~\cite{RAYVDM} generators for 
 $\gamma p$ interactions. With these models  large samples of photoproduction
 events were generated which contained all  classes of events (elastic,
 soft hadronic collisions, hard scattering
processes  and heavy flavour  production).
% Samples of events with an electron candidate in the BEMC, as defined by the
% analysis criteria presented in the next section, were selected 
%  which correspond to about 3 times the
% luminosity of the data.          

 It was found  that  
 about 10\% of the DIS  data at HERA consists of events with a large  gap 
 in pseudo-rapidity around the proton remnant direction
\cite{diffrac}. These events
 were found to be compatible with diffractive exchange and are
 well described  by the model RAPGAP~\cite{rapgap} as 
deep-inelastic scattering on a colourless object 
--termed a pomeron--
emitted from the proton. The hadronic final state 
of these events is also
well described by RAPGAP which includes ARIADNE for QCD effects. The
RAPGAP Monte Carlo simulation 
 was used to check the effect of the large rapidity gap events
on the vertex reconstruction efficiency
%, and the results were included in the systematics
%on the Monte Carlo description of the hadronic final state.
which depends
 mostly on  the final state topology of the events. 
 Differences between rapidity gap and ``standard'' DIS events
of up to $2\%$ were found 
at large $y>0.4$ and smaller at low $y$,
  and were included in the systematic error of $F_2$.                           

 For 
 the events generated with the models described above the detector response 
was simulated in detail~\cite{h1detec} using a  program based on  
  GEANT~\cite{GEANT}.
  The simulated Monte Carlo events were subjected to the same
 reconstruction and analysis chain as the real data.

%-------------------------------------------------------------------- 

\section{Event Selection}  
%-------------------------------------------------------------------- 
 
%The acceptance  for
%scattered electrons in the BEMC and the Liquid Argon Calorimeters        
%are indicated. The extension of the kinematic region by shifting the vertex  
%by 62 cm is also shown.}          
       
The low $Q^2$ DIS events in the backward region 
were triggered by an energy cluster in the BEMC ($E_e' > 4$ GeV)
which was
not vetoed by the TOF. The high $Q^2$ events
were triggered by  requiring an electromagnetic energy cluster 
 in the LAr calorimeter ($E_e' > 8 $ GeV).
A trigger of 
lower energy threshold ($E_e'> 6$ GeV)  also accepted  the event if
there was simultaneously a tracking trigger. In the region of the final 
$F_2$ data presented below 
the trigger efficiency, which  has
been determined from the data,  is about $80\%$ for 
$E'_e \sim 8$ GeV, and becomes 
larger than 
$99\%$ for  $E'_e > 10$ GeV. %(non-radiative DIS sample)
%  and larger than $80\%$ for 
%$E'_e \sim 8$ GeV. % (tagged radiative DIS sample).

\subsection{Selection of Deep-Inelastic Scattering Events} 
         
Deep-inelastic scattering events in H1 are  identified  by the
detection  of the scattered electron in the BEMC or LAr calorimeter
and the presence of a reconstructed interaction vertex.
The electron identification cuts, fiducial volume and vertex requirement
are detailed in Table 1.

These selection criteria
 follow closely those of the 1993 data analysis~\cite{H194}.
%For the "open triangle'' sample (see section 1) the angular cut could be 
%relaxed from 173$^0$ to 174$^0$.
For the low $Q^2$ nominal vertex sample ($Q^2 \le $ 120 GeV$^2$)
an additional cut $r_{BPC} < 64$ cm  is
applied, where $r_{BPC}$ is the radial distance of the electron hit  in the BPC to the beam axis. 
This cut prevents the electron from  entering  the transition region between
the BEMC and the LAr calorimeter where the energy corrections are large and
depend strongly on the impact point. 
For the same reason, the high $Q^2$ events ($Q^2 > $ 120 GeV$^2$)  
are accepted only if the electron cluster is fully contained in the LAr
calorimeter. Despite these conditions, the measurement 
could also be performed  for intermediate $Q^2$ ($Q^2\sim 120$ GeV$^2$)
due to the  $\pm 30$ cm
spread  of the  event vertex position around its nominal position.

%The electron identification efficiency as determined from
%Monte Carlo simulation  is better than $97\%$.
% apart
%from the BEMC-LAr transition region where it amounts to approximately
%$92\%$.

The scattered electron is identified with the  electromagnetic
cluster of maximum  energy  which satisfies the 
estimator cuts of Table 1. 
The electron identification efficiency, determined from
Monte Carlo simulation studies,  is better than $97\%$ except at % low $Q^2$
  $Q^2 \le$ 6.5 GeV$^2$ where it falls to 94\% at the lowest
  $x$ values.

%If two electron candidates are present in the same event
%the one with the highest energy is selected. This introduces a 
%misidentification probability of at most 3\% at the smallest electron
%energy which was considered in the systematic error calculation.

%------ 
%       
\begin{table} [htb] \centering   
\begin{tabular}{|c|c|c|c|c|} 
\hline  
   &  {}{low $Q^2$ \small{(shifted vtx)}}           
   &  {}{low $Q^2$ \small{(nominal vtx)}}           
   &  \multicolumn{2}{c|}{high $Q^2$}\\   
\hline  
         &  E, $\Sigma$ method           
         &  E,     $\Sigma$ method           
         &  E method    & $\Sigma$ method  \\       
\hline  
 $\theta_e/^o$          
         &  $\leq 176  $   
         &  $ < 173  $   
         &  $ < 150  $   &  $\leq 153  $   \\      
 $E'_e$/GeV       &  $> 11   $   
         &  $> 11   $   
         &  $> 11   $   &  $> 11   $   \\           
 $z_{vertex}/$cm  &  $ 67 \pm$ 30
         &  $ 5 \pm$ 30 
         &  $ 5 \pm$ 30 &  $ 5 \pm$ 30  \\          
 electron   identif.      &  $\epsilon_1 < 5$~cm  
         &  $\epsilon_1 < 5$~cm  
         & $\epsilon_3> 50\%$   & $ \epsilon_3> 65\%$\\      
 electron identif.         & $\epsilon_2 < 5$~cm   
         & $\epsilon_2 < 5$~cm 
         & $\epsilon_4 >3\%$  & $\epsilon_5 < 30$ mrad\\     
\hline  
% events  &  9000       & 12000  
%         & 165000      &  221000
%         & 6000        &  7000     \\     
%\hline 
\end{tabular}           
\caption  {\label{taevs1}        
\sl Summary of event selection criteria for the shifted %vertex data   
    and the nominal vertex (vtx) data at low and high     
    Q$^2$. The approximate event numbers are 10000, 220000 and 9000 events
    respectively.
For the open triangle data sample the $\theta_e$ cut is
174$^{\circ}$.
For the electron identification several estimators were used:        
    $\epsilon_1$: electron cluster radius;
    $\epsilon_2$: smallest distance from the closest hit in the BPC to the     
         centroid of the electron cluster;           
    $\epsilon_3$: fraction of the electron  energy deposited in the 4 
         most energetic cells of the cluster;     
    $\epsilon_4$: fraction of the electron energy deposited in the first       
         3 radiation lengths of the calorimeter;    
    $\epsilon_5$: 
%angular distance  between the electron cluster and the       
%         associated track.
   angle between the line connecting the vertex to the centroid of the electron
   cluster and the associated track.}      
\end{table}

At low $Q^2$  the main sources of non-$ep$ background are due to      
proton beam interactions with residual gas and beam line elements     
upstream of the H1 detector.     
At high $Q^2$ the main background is due to cosmic ray events         
and muons travelling off axis parallel to the proton beam.   
An  efficient reduction of these background contributions is       
provided by the minimum energy and the vertex 
requirements discussed above.  
The number of residual beam-induced background events       
was  estimated from non-colliding bunch studies,
 and the number of cosmic events from scanning.
Both together represent less than 1\% of the  number of      
selected events in any ($x,Q^2$) bin.

%At high $Q^2$ the main background is due to cosmic ray events         
%and muons traveling off axis parallel to the proton beam.   
%Requiring a reconstructed vertex rejects most of them.       
%The remaining background was estimated to be smaller than $1\%$.      
        
The only significant background to DIS from $ep$ interactions         
is due to photoproduction events where the scattered        
electron escapes the detector along the beam pipe but in which        
an energy cluster from the hadronic final state fakes a scattered
electron.  
About 10\% of these events  are identified  as photoproduction     
background if the scattered electron is found in the electron tagger. 
 Photoproduction events were simulated to estimate this background.
The photoproduction background was subtracted statistically bin by bin.       
Only 12 bins, out of a total of 193 ($x,Q^2$) bins,     
    have a contamination larger than $3\%$. This contamination never  
exceeds $15\%$ in any bin.

\begin{figure}[htbp]    \unitlength 1mm
\begin{center}
\begin{picture}(120,85)
\put(-11,-3){
\begin{picture}(0,0) \put(50,80){({\large \bf a})} \end{picture}
\begin{picture}(0,0) \put(135,80){({\large \bf b})} \end{picture}
\epsfig{file=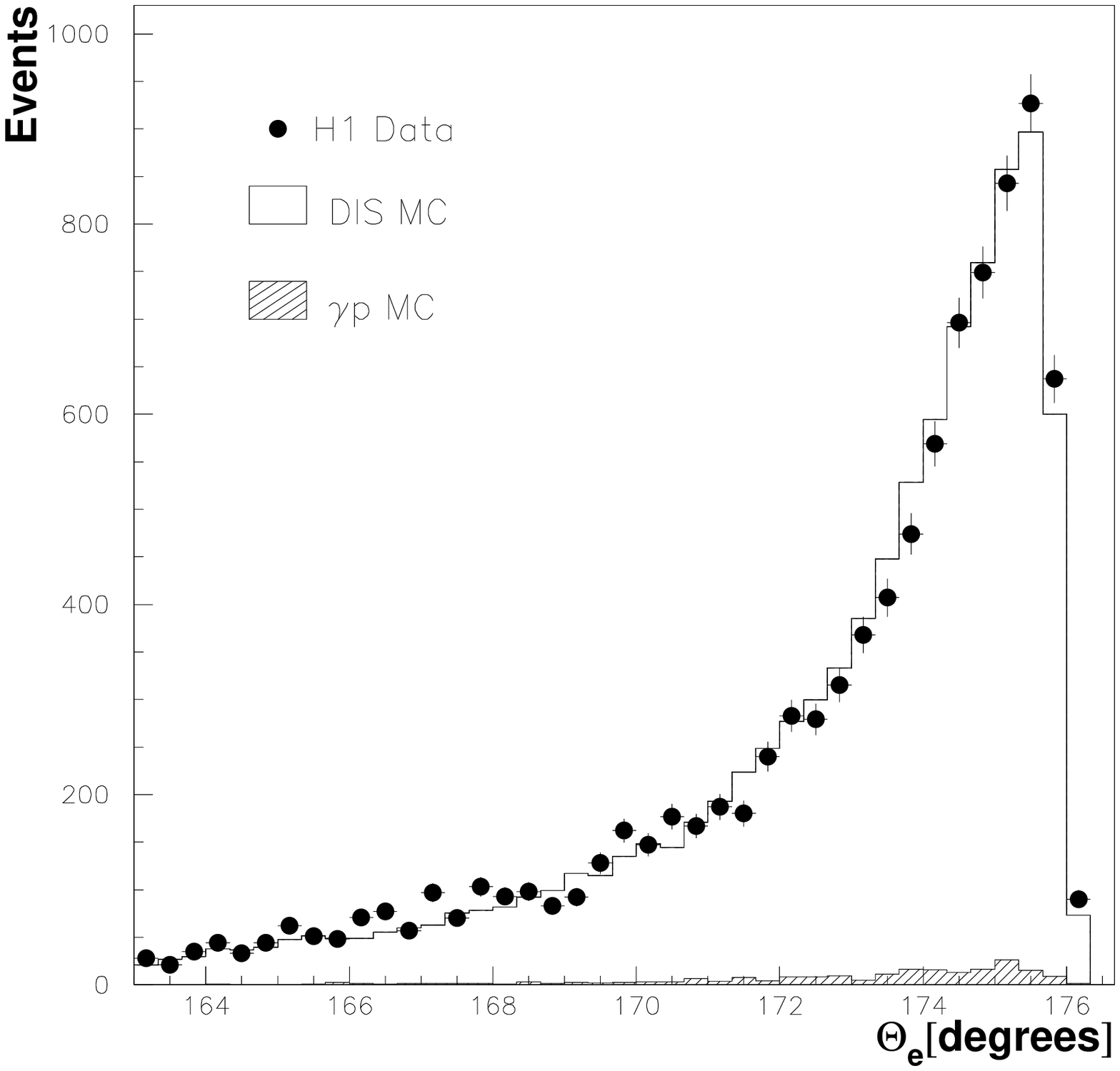,width=7.3cm,bbllx=80pt,bblly=70pt,bburx=560pt,bbury=700pt}
\epsfig{file=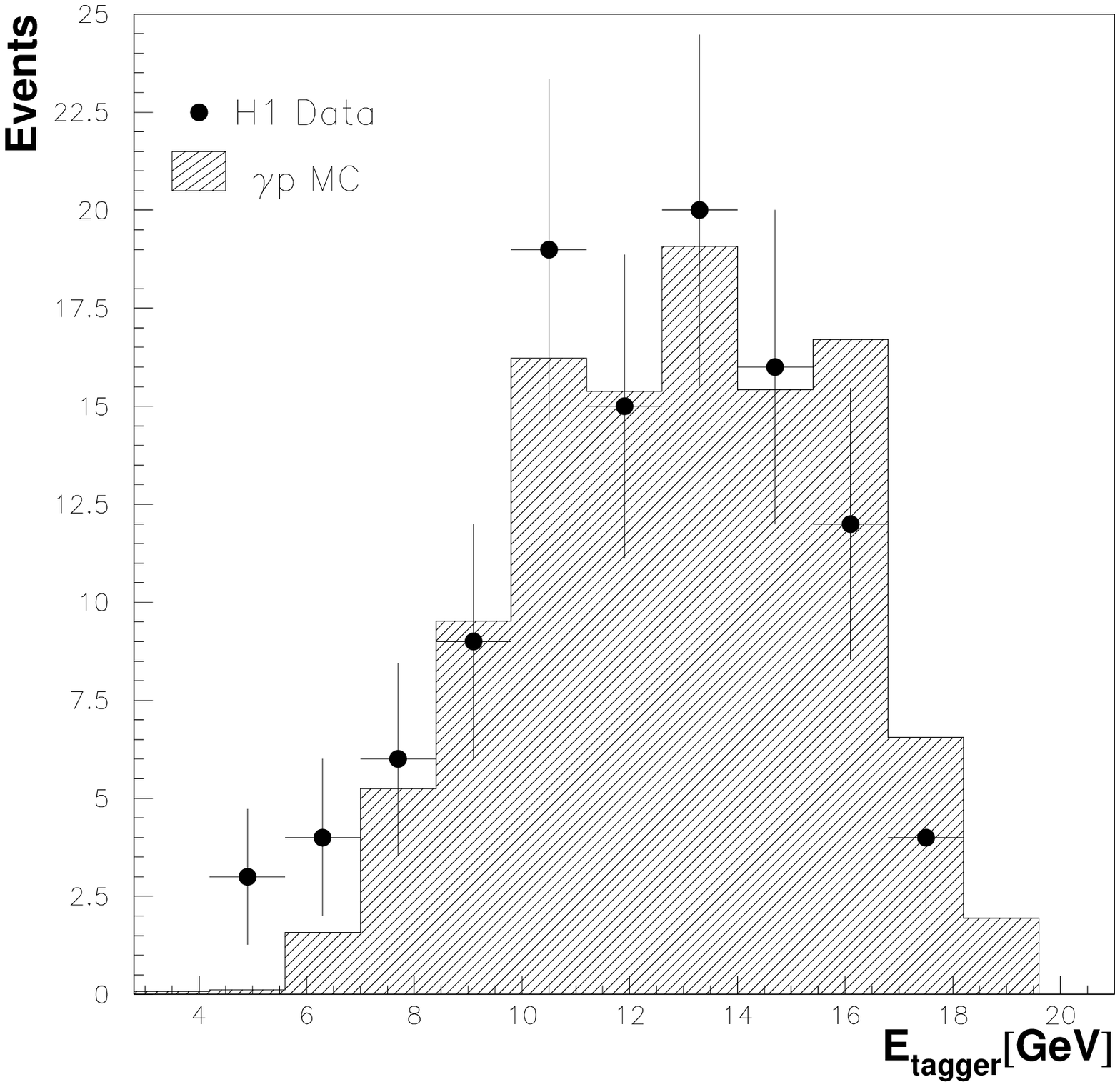,width=7.3cm,bbllx=10pt,bblly=70pt,bburx=490pt,bbury=700pt}   
%\put(50,80){({\large \bf a})}
%\put(135,80){({\large \bf b})}
}
\end{picture}
\end{center}   
\caption[]{\label{eplotsvx}      
\sl Shifted vertex data:  experimental and Monte Carlo  
distributions of a) the polar angle  of the scattered 
electron  and b)
  the energy of the scattered electron 
in photoproduction background events detected in the 
electron tagger.  }
\end{figure}   
        
Figure \ref{eplotsvx}a shows the distribution of the angle
of the scattered electron for the shifted vertex data compared to 
the Monte Carlo simulation weighted with the
measured structure function (see section 6). 
The Monte Carlo simulation is normalized
to the luminosity  and agrees well with the 
data illustrating the level of residual  background  %to which the background is 
 in the low $Q^2$ sample.
%The same procedure is also used in Fig.~3a, 4a, 4b presented hereafter. 
%The Monte Carlo gives a good description of the data.
%The simulation describes the data well.
% The contribution of the photoproduction background
%determined from Monte Carlo calculations, is also indicated.
%, and is only
%significant at low energies. 
%The Monte Carlo simulation for the 
%photoproduction background
%are controlled by a subsample  ($\sim 10\%$)
%of events for which the scattered electron
%is detected by the small angle electron tagger of the luminosity system.
In Fig.~\ref{eplotsvx}b the normalized   energy spectrum in the electron
tagger is shown which is reproduced by the background photoproduction 
event simulation.

\begin{figure}[htbp]    \unitlength 1mm
\begin{center} 
\begin{picture}(120,85)
\put(10,-1){         
\begin{picture}(0,0) \put(-10,72){({\large \bf a})} \end{picture} 
\begin{picture}(0,0) \put(122,72){({\large \bf b})} \end{picture}
\epsfig{file=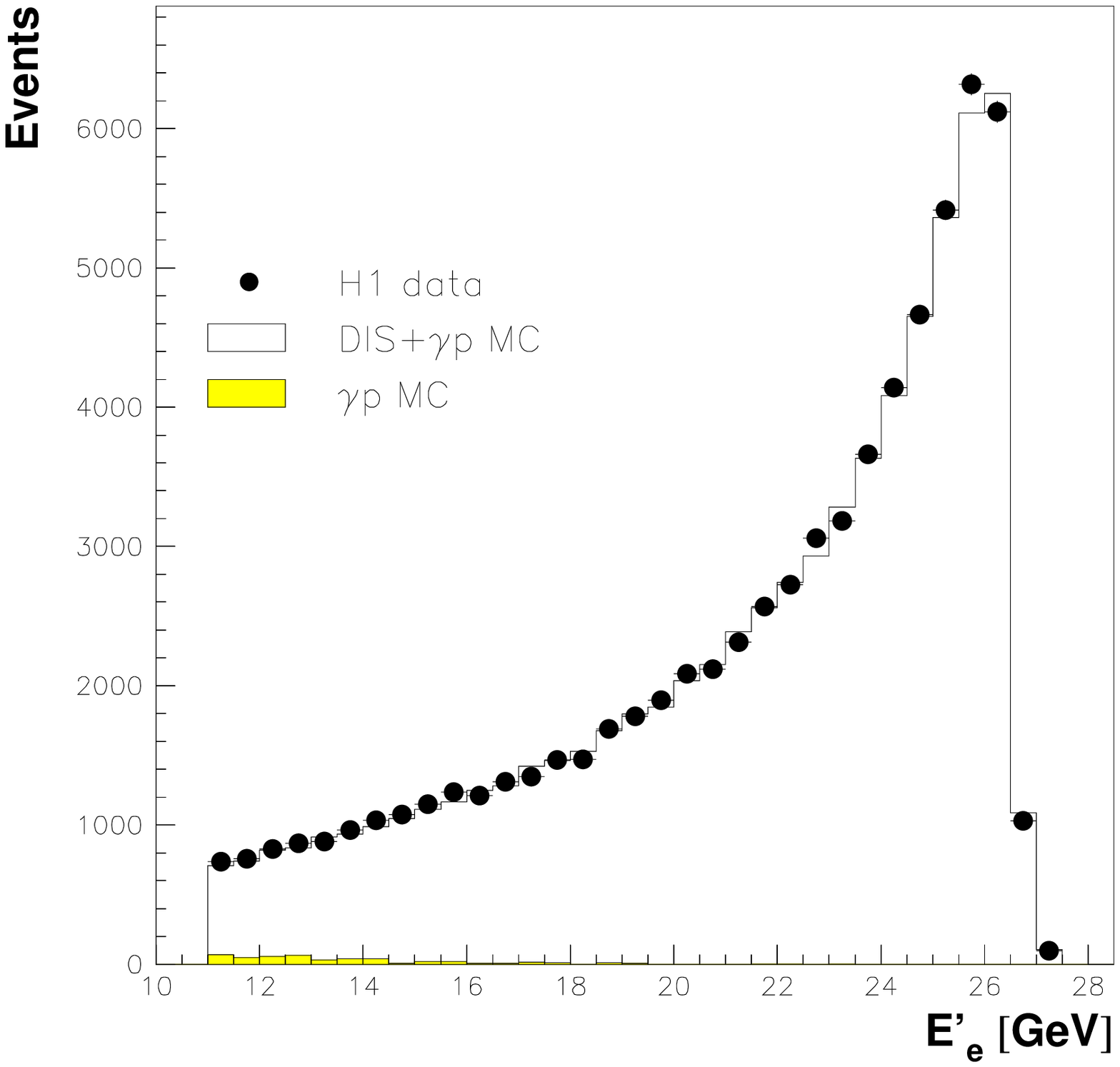,width=6.5cm,bbllx=250pt,bblly=150pt,bburx=650pt,bbury=550pt}
\epsfig{file=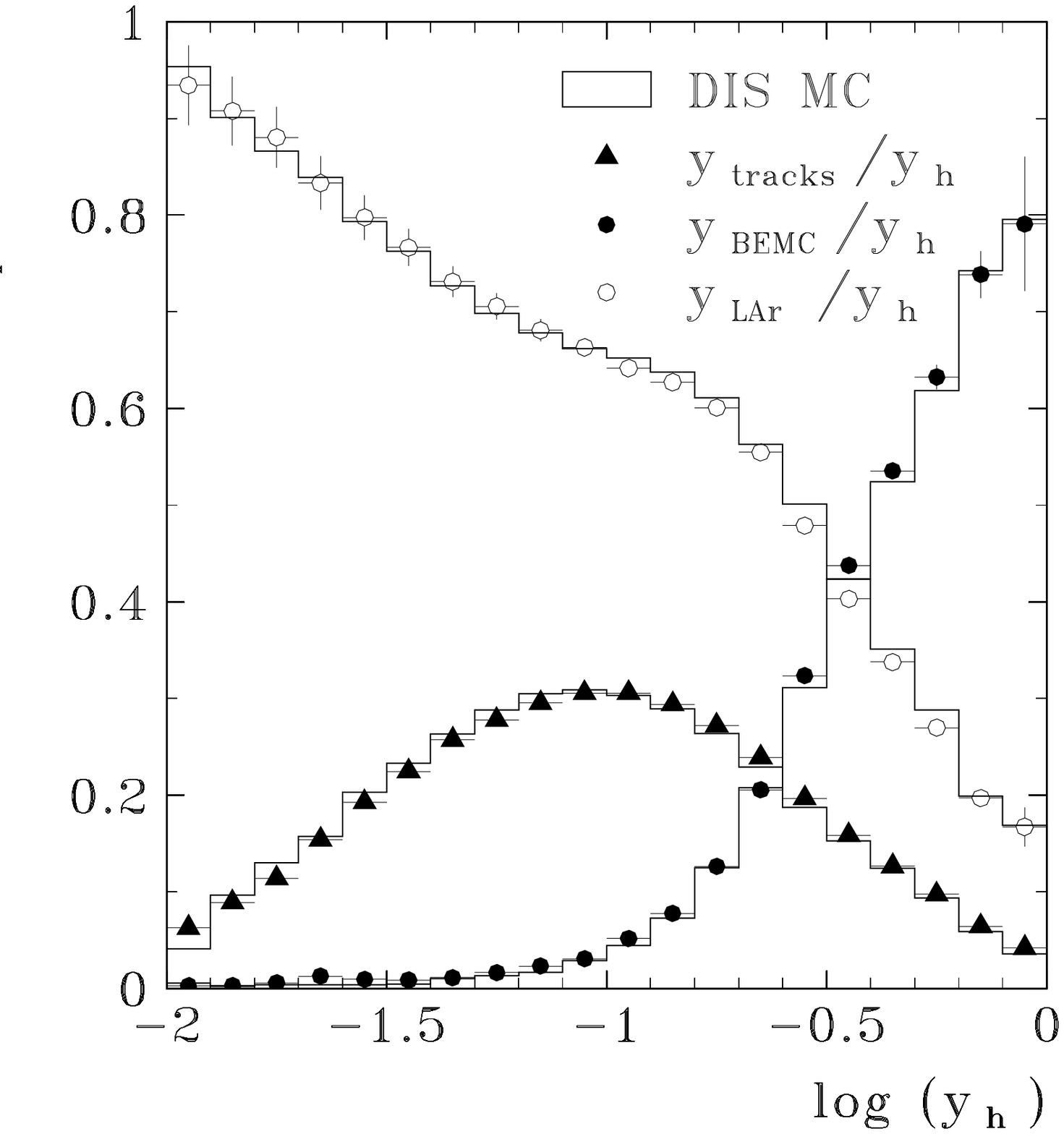,width=6.5cm,angle=90,bbllx=30pt,bblly=530pt,bburx=430pt,bbury=730pt} 
}
\end{picture}  
\end{center}   
\caption[]{\label{eplotnvx}   
\sl Nominal vertex data with the
 scattered electron in the BEMC ($Q^2 \leq$ 120 GeV$^2$): experimental
 and Monte Carlo distributions a) of the scattered  electron
energy  and
b) of the  fraction of $y_h$ contributed
 by the tracks, the LAr calorimeter and the BEMC.}
\end{figure}
\begin{figure}[h!]    
\begin{center}          
\begin{picture}(0,0) \put(35,72){({\large \bf a})} \end{picture} 
\begin{picture}(0,0) \put(115,72){({\large \bf b})} \end{picture}
\epsfig{file=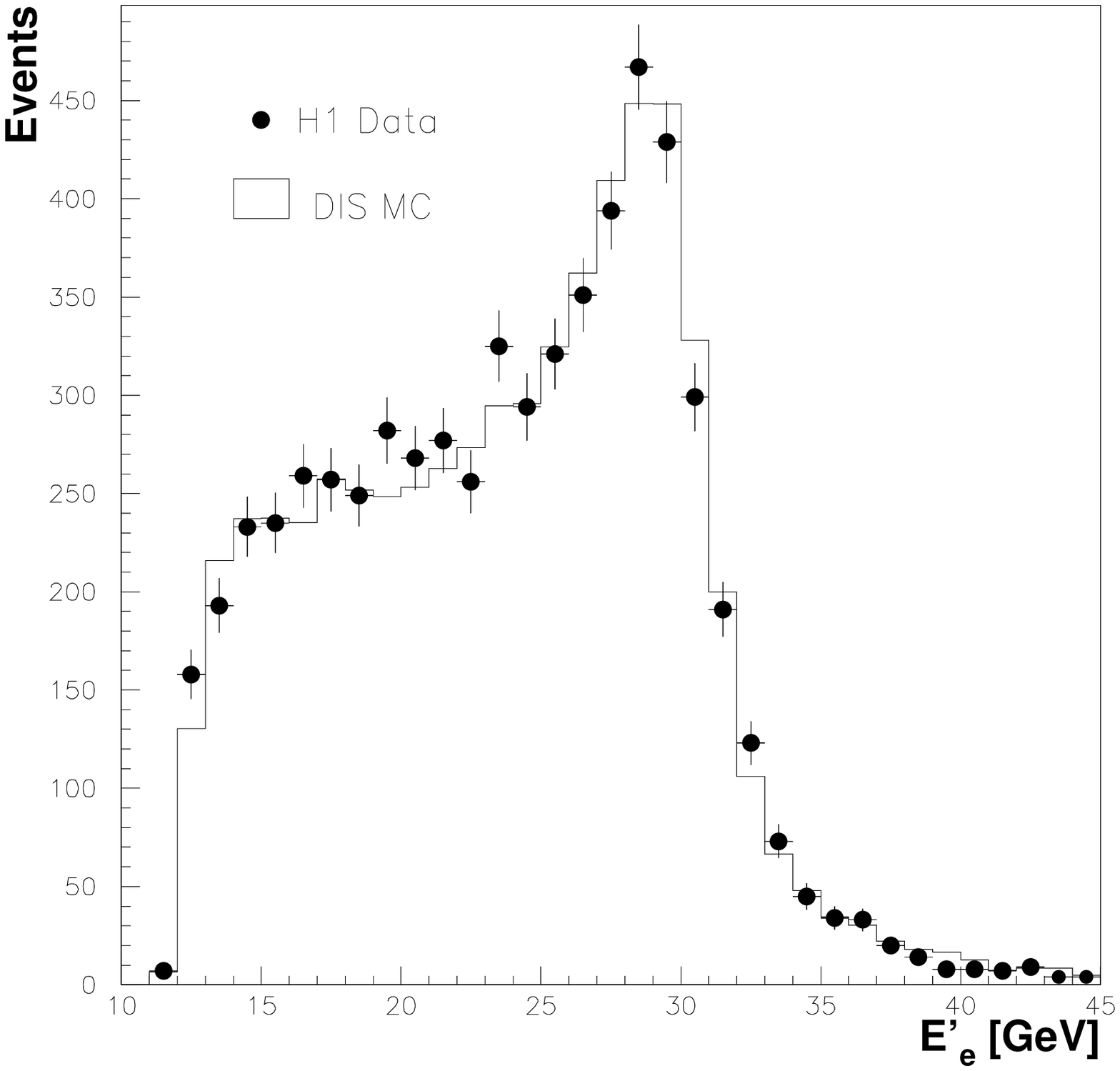,width=6cm,bbllx=250pt,bblly=100pt,bburx=650pt,bbury=800pt}   
\epsfig{file=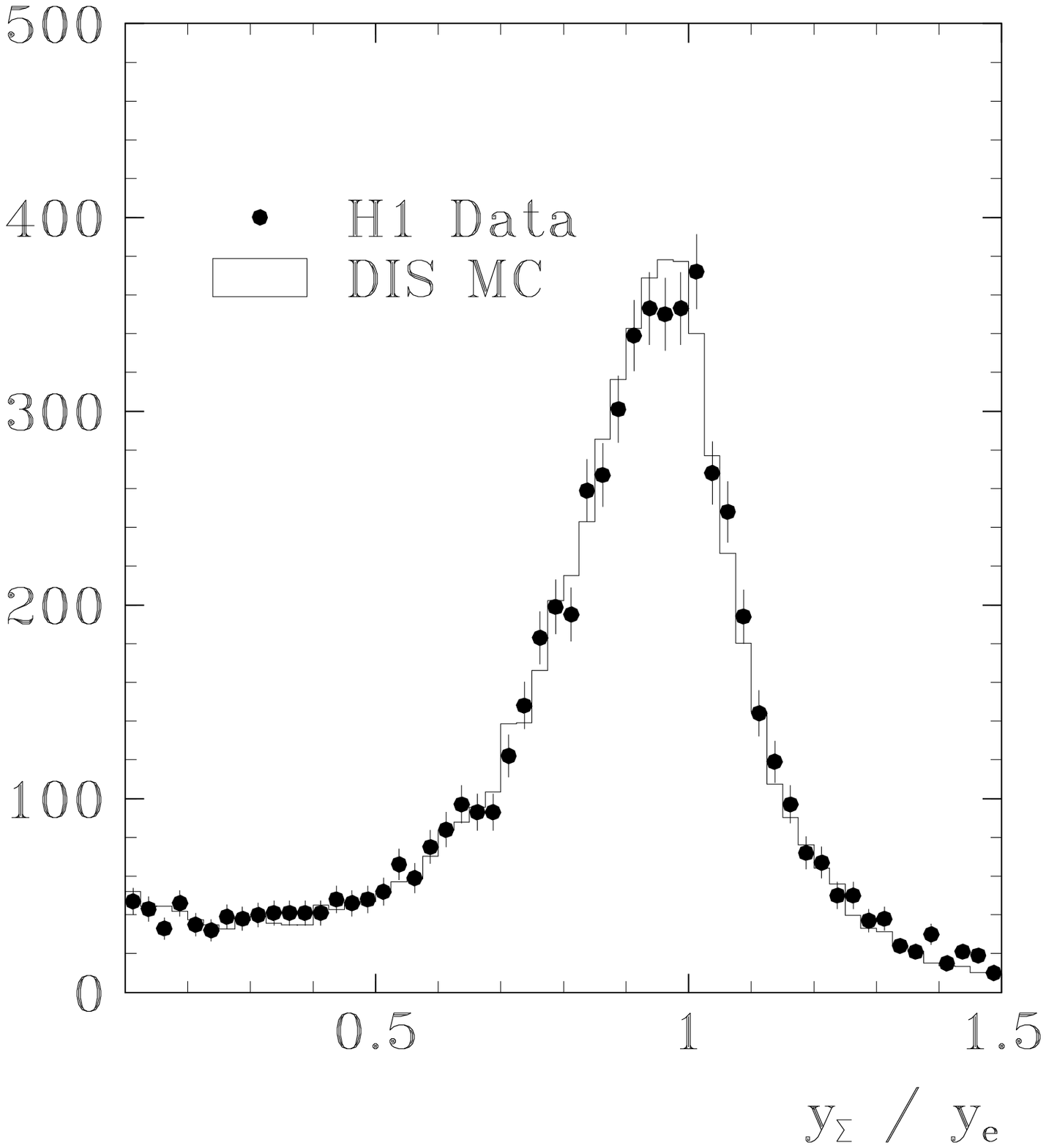,width=6cm,bbllx=0pt,bblly=550pt,bburx=365pt,bbury=725pt,angle=90}    
\end{center}   
\caption[]{\label{eplothiq2}   
\sl High $Q^2$ data:  experimental and Monte Carlo distributions
 a) of the energy of the scattered electron detected 
in the LAr calorimeter and b) of the ratio
  ${y_{\Sigma}}/{y_e}$, for $y_e \ge$ 0.05.}
\end{figure}   
    
Figure \ref{eplotnvx}a shows the distribution of the energy
of the scattered electron for the high statistics
nominal vertex data.
The simulation gives an excellent  description of the data 
from the low energy % kinematically corresponds to 
up to the   so called kinematic peak region, i.e. the region
around the value of the incident
electron beam energy. This agreement was achieved after a spatially
dependent calibration 
of the data and  Monte Carlo response~\cite{panitch} using the 
double-angle method~\cite{hoeger}.
The small remaining
contribution of the photoproduction background is also shown.
%Figure \ref{eplotnvx}b illustrates  on the same data sample how $y_h$
%(from which $y_{\Sigma}$ can be derived  $y_{\Sigma}=y_h/(1+y_h-y_e)$ and
%$y_h=\Sigma/2E_e$) is measured
%by using a combination of central tracks and  calorimeter
%cells~\cite{pineiro}. 
In Fig.~\ref{eplotnvx}b the fractions of $y_h$ originating from  tracks,  BEMC and  LAr
 calorimeter  are given as a function of $\log_{10} y_h$. 
In this analysis the $y_h$ variable is determined by using a combination of central tracks and  calorimeter
cells~\cite{pineiro}.
An isolation criterion is used to avoid counting
the energy of the LAr cells originating from a track already used in $y_h$.
For $y < $0.15, i.e. in the region where the $\Sigma$ method will be
used for the  $F_2$ result,  
the measurement
is dominated by the track
reconstruction and the LAr measurement 
(Fig.~\ref{eplotnvx}b). At larger $y$,
the BEMC contribution
plays an increasing role due to the low energy particles which
accumulate in the backward direction. 
The DIS Monte Carlo simulation describes well
 these fractions
in the complete kinematic range.

Figure \ref{eplothiq2}a  shows the distribution of the energy
of the scattered electron  detected
in the LAr calorimeter. It is well described by 
 the Monte Carlo simulation.
A detailed calibration was carried
out  by comparing events from the kinematic peak 
at low $y$ ($< 0.1$) with simulation, including corrections for
 the  energy lost due
to the dead material  between the wheels which make up the LAr
calorimeter~\cite{lipinski}. 
This procedure  has been  cross checked with the 
double-angle method. 
%both on data and Monte Carlo resulting in the good agreement.

Figure \ref{eplothiq2}b shows the ratio $y_{\Sigma}/y_e$ in the 
high $Q^2$ sample   compared to the Monte Carlo 
expectation. The resolution of this ratio, which is 
 calculated for $y_e > $ 0.05,
and thus of $y_{\Sigma}$ is  better than 13\% in this kinematic region.
The ``tail" visible at values below 0.7
is due to  radiative events,  and 
is well described by the Monte Carlo simulation.
%These events account for  half of the 2.5\% deviation of the ratio
% from the expected value of 1.        
     
%       

\subsection{Selection of Deep-Inelastic Radiative Events}   

A sample of deep-inelastic events   with an
energetic photon ($E_{\gamma} > 4$ GeV) emitted collinear with the incident 
electron was selected. 
 These radiative events can be 
interpreted as %non-radiative 
deep-inelastic scattering
events with a reduced  (``true") incident energy 
$E_t = E_e-E_{\gamma}$ % $~\cite{krasny},
which can be reconstructed due to the additional detection of the 
radiated photon in the small angle photon tagger of the luminosity 
system.
When using the E method,  the %true 
kinematic variables  $y_t$ and
$Q_t^2$ %for such an $ep$ collision 
are
obtained by replacing in eq.\ref{kinematics1} 
 the nominal beam energy 
by the reduced energy $E_t$.
%
%\begin{equation}
%  y_t   =1-\frac{E'_e}{E_t} \sin^{2}\frac {\theta_e} {2}
%   \hspace*{2cm}
%   Q^2_t = \frac{E^{'2}_e \sin^2{\theta_e}}{ 1-y_t}
%Q^2_t = 4E'_eE_t\cos^{2}\frac{\theta_e} {2}
%\end{equation}
% demonstrating  the access to  smaller $Q_t^2$ values for given
%values of $E_e'$ and $\theta_e$.
Note that $Q^2_{\Sigma}$ and $y_{\Sigma}$ are unchanged by the 
$E_e \rightarrow E_t$ transformation while $x_{\Sigma}$ is affected.

A first experimental study of this process at HERA
has been  published~\cite{h1radpap,favart}
 by the H1 
collaboration 
using  1993 data, which where however too limited 
in statistics for a quantitative study of the proton structure.
%of the proton structure function $F_2(x,Q^2)$. 
The larger  
integrated luminosity of the 1994 data
%(2.9 pb$^{-1}$)
permits    a significant $F_2$ measurement 
for $Q^2$ values down to $1.5\, {\rm GeV}^2$. 
The ZEUS collaboration~\cite{zeuslowq2} recently published results 
on $F_2$ using this method.

\begin{figure}[tb!]    \unitlength 1mm
\begin{center} 
\begin{picture}(120,88)
\put(0,-1){              
\begin{picture}(0,0) \put(45,80){({\large \bf a})} \end{picture} 
\begin{picture}(0,0) \put(45,40){({\large \bf b})} \end{picture}
\begin{picture}(0,0) \put(130,80){({\large \bf c})} \end{picture}
\epsfig{file=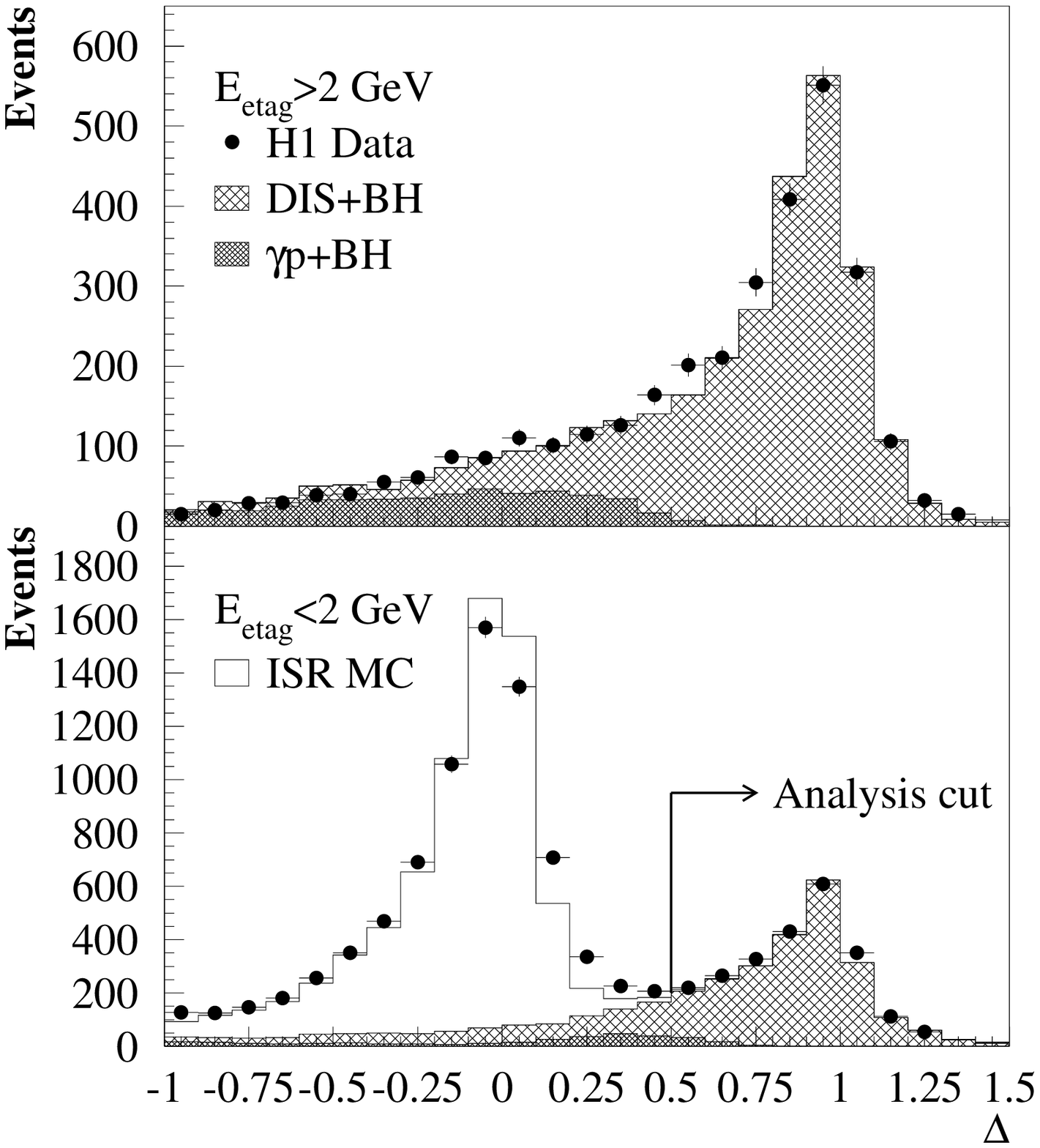,width=6cm,bbllx=200pt,bblly=0pt,bburx=550pt,bbury=450pt}
\epsfig{file=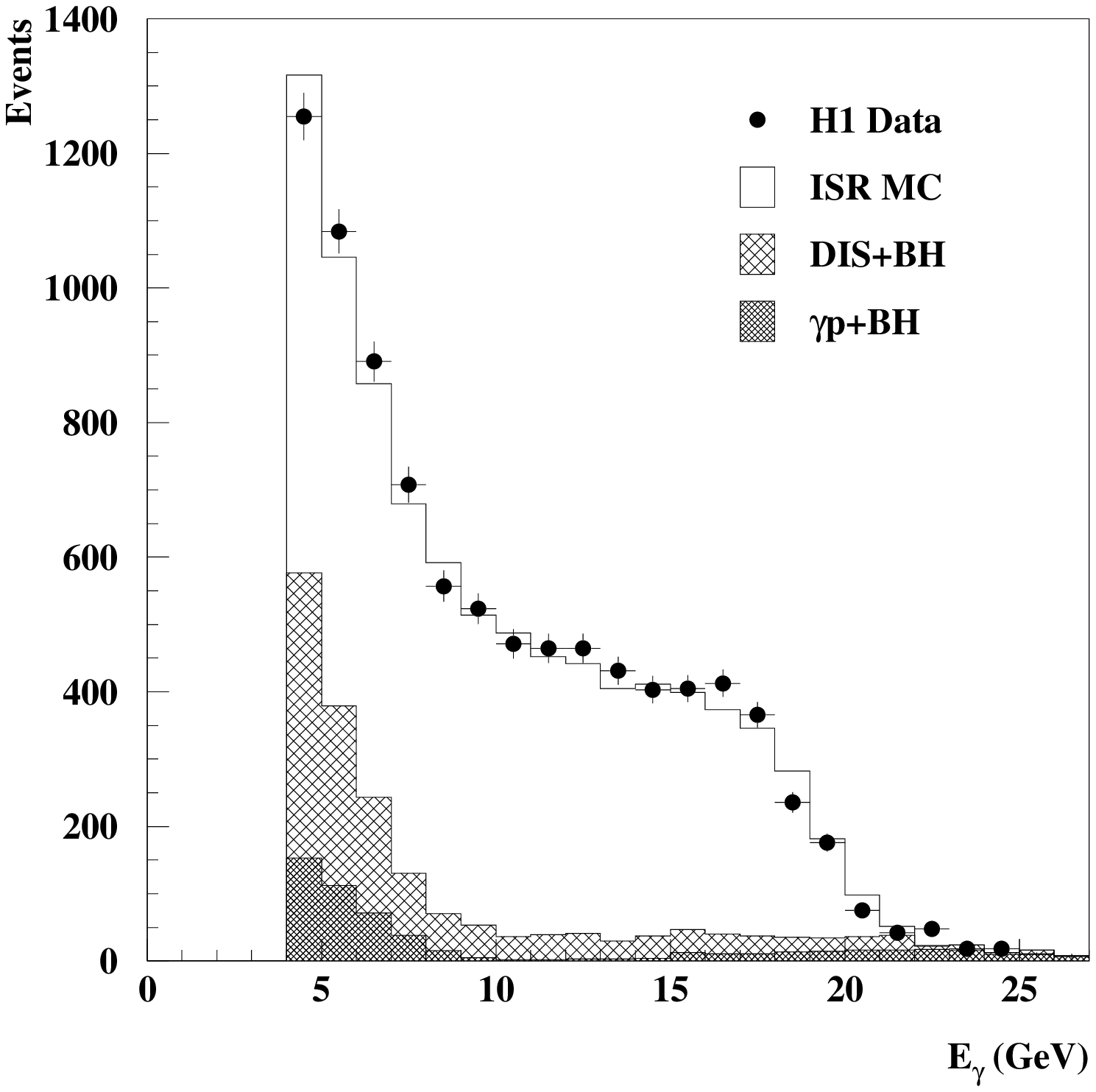,width=6cm,bbllx=100pt,bblly=0pt,bburx=400pt,bbury=350pt}
}
\end{picture}
\end{center}   
\caption[]{\label{eplotrad}   
%\sl Radiative events:  Experimental and Monte Carlo distributions
% of $\Delta$ (eq.4) with a) the energy detected in
%the electron tagger ($E_{etag}$) bigger than 2 GeV;
% b) with $E_{etag} <$ 2 GeV
%and c)  of the photon energy detected in the photon tagger.
%The analysis cut in b) indicates that the region of $\Delta >$ 0.5 was 
%excluded from the analysis.}
\sl Radiative events:  experimental and Monte Carlo distributions
 of $\Delta$ (eq.~6) with a) the energy detected in
the electron tagger ($E_{etag}$) bigger than 2 GeV;
 b) with $E_{etag} <$ 2 GeV
and c)  distribution   of the photon energy detected in the photon tagger.
The analysis cut in b) indicates the region of $\Delta >$ 0.5
excluded from the analysis.
The full solid line in b) and c) 
represents the sum of all three contributions in the 
Monte Carlo:  DIS initial state radiation events (ISR MC),  DIS events with
a BH overlap  (DIS + BH) and photoproduction events with a BH overlap 
($\gamma p$ + BH). }
\end{figure}

 A summary of the selection criteria  
of the final sample of about 8200 events is given in 
Table \ref{taevs2}~\cite{H1RAD}.  
The event     
selection for radiative 
events is similar to 
the one  for low $Q^2$ non-radiative events, apart from the
additional requirement  of 
 a detected photon with at least 4 GeV in the
small angle photon tagger of the luminosity system.
This requirement also  reduces the photoproduction background.
 Therefore the minimum scattered electron energy 
   can be lowered to 8 GeV. 
%efficiently rejected by comparing the energy measured in the photon 
%detector with the expected one derived from the kinematic quantities 
%measured in the main detector.

The selected sample 
 contains both radiative DIS events  
and  pile--up events due to overlaps of DIS 
and $\gamma p$ events with Bethe Heitler events 
%in the same same bunch crossing  
 in a time window
of $\pm5$~ns. The pile-up events
are partly removed from the sample by   
requiring the energy in the electron tagger, $E_{etag}$, to be  
less than 2 GeV, 
but the majority  of them remains.
%The majority of these background events  
%has the bremsstrahlung electrons outside the acceptance region of the        \bibitem{qcdbcd} M. Virchaux and A. Milsztajn, Phys.\ Lett.\  {\bf B274} (1992) 221.

%electron tagger.    
  The  background  can be controlled through the redundancy
of the 
true electron beam energy measurement $E_t$.   
 For radiative DIS 
events we expect measurements of the quantity    %$\Delta$
  \begin{equation}     
  \Delta \equiv [E_{\gamma} - E_e(y_e-y_h)]/E_{\gamma}\label{kinematics3}     
  \end{equation}       
to be concentrated around zero while 
for  pile-up DIS events a concentration  around one is expected.
Here $y_e$ and $y_h$ are calculated according to eqs.~\ref{kinematics1} and
\ref{kine4}. 
%Here $
%  y_h = \Sigma \, (E_h - p_h \cos \theta_h)/2 E_e $, where the sum
%is over all final state hadrons.
The distribution of $\Delta$ is shown in Fig.~\ref{eplotrad}
for a sample of events with a) $E_{etag} > 2 $ GeV (dominantly 
$ep$ collisions with BH overlap events) and b) $E_{etag} < 2 $ GeV.
The data are compared with Monte Carlo simulation.
The pile-up sample in Fig.~\ref{eplotrad}a shows a clear peak for 
$\Delta = 1$, and is well described by the sum of 
simulated DIS and $\gamma p$ 
distributions with overlap of BH events.
Fig.~\ref{eplotrad}b shows a  peak for $\Delta = 1$ from residual
pile-up events for which the electron from the BH event was not
detected, and a peak around $\Delta = 0$ from genuine radiative events.
Radiative events are  selected 
in this analysis by requiring $\Delta < 0.5$. The  background of 
pile-up events as estimated by the Monte Carlo
simulation studies is  subtracted statistically.
The  remaining background from overlap $\gamma p$ and DIS events
estimated from Monte 
Carlo studies amounts to 8\%, with at most 15\% in an $x,Q^2$ bin. 
In Fig.~\ref{eplotrad}c the photon energy spectrum as measured in the 
photon tagger is shown for the selected sample and compared with simulated
signal and background distributions. There is a good agreement 
between data and Monte Carlo simulation.

\begin{table} [htb!] \centering   
\begin{tabular}{|c|c|} 
\hline  
  &  {low $Q^2$ (\small{radiative events})}\\
\hline  
         &  E, $\Sigma$ method      \\       
\hline  
 $\theta_e/^o$          
         &  $\leq 174  $  \\     
 $E'_e$/GeV       &  $> 8   $      \\           
 $z_{vertex}/$cm  &  $ 5 \pm$ 35 \\         
 electron   identif.      &  $\epsilon_1 < 5$~cm   \\  
 electron   identif.         & $\epsilon_2 < 4$~cm    \\  
 $E_{\gamma}$/GeV       & $> 4$   \\
 $E_{etag}$/GeV         & $< 2 $ \\
 $\Delta$  & $< 0.5$\\
%\hline 
%         events  &  5000 \\   
\hline  
\end{tabular}           
\caption  {\label{taevs2}        
\sl Summary of event selection criteria for the  radiative event sample.   
For the electron identification two estimators were used:        
    $\epsilon_1$: electron cluster radius and
    $\epsilon_2$: smallest distance from the closest track to the     
         centroid of the electron cluster.
The variable $\Delta$ is defined in eq.~\ref{kinematics3}. }
\end{table}

\section{Structure Function Measurement} 
The  structure function $F_2(x,Q^2)$     
was derived   
from the one-photon exchange cross section   
\begin{equation}       
  \frac{d^2\sigma}{dx dQ^2} =\frac{2\pi\alpha^2}{Q^4x}      
    (2-2y+\frac{y^2}{1+R}) F_2(x,Q^2).   
\label{dsigma}         
\end{equation}         
The structure function ratio $R=F_2/2xF_1 - 1 $    
has  not yet been measured  at HERA. It    
was calculated using the  QCD relation~\cite{Altmar}
with the NLO strong coupling constant~\cite{marciano}  and  the        
GRV structure function parametrization.    
% Here $M_p$ is the proton mass.
%At small $x$ and $Q^2$   
%the assumed $R$ values which are given in
%Tables~\ref{f2tabl},~\ref{f2tabh} can be as large as 1. 
Note that 
a 20\% error on $R$ corresponds to about
2\% uncertainty on $F_2$ at $y=0.6$ for $R$ of about  0.6. 
 The  $R$ values are quoted in Tables~\ref{f2tabl} and \ref{f2tabh};
 no extra effect of the R uncertainty on $F_2$ was considered.
        
Compared to the previous H1 analysis~\cite{H194}    
the $F_2$ measurement has been extended to lower and         
higher $Q^2$ (from $4.5-1600$~GeV$^2$ to  $1.5-5000$~GeV$^2$),        
and to lower and higher $x$ (from $1.8 \cdot 10^{-4}\le x\le 0.13$ to         
$3 \cdot 10^{-5}\le x \le 0.32$).         
The determination of the structure function requires         
the measured event numbers to be converted
to the bin averaged cross section         
based on the Monte Carlo acceptance calculation. The binning in $x$ was 
governed by the detector resolution and could be chosen to be rather fine 
since the E and $\Sigma$ methods were used in the optimum range at low 
and high $x$, respectively. The $x$ resolution is better than 20\%.
The $Q^2$ resolution is about 5\% and the 
number of bins in $Q^2$  was adapted to statistics.   
 All detector efficiencies were determined from the data     
utilizing the redundancy of the apparatus. Apart from very small      
extra corrections, %($\leq$3\%) 
all efficiencies were correctly    
reproduced by the Monte Carlo simulation. 
The bin averaged cross section was        
corrected for higher order QED radiative contributions       
using the  program HECTOR \cite{HEC}.      
Effects due to $Z$ boson exchange 
at present values of $Q^2$ and $y$ are smaller than 3\% and were treated    
as part of the radiative corrections.     
        
Different data sets are available  %to this analysis          
%     : nominal vertex data,      
%"open triangle" data ,  "shifted vertex" data and "satellite'' data,  
which, for a given ($Q^2,x$) interval,  use different parts of the detectors. 
Thus many cross         
checks could be made in kinematic regions of overlap
for the two kinematic      
reconstruction  methods and these gave        
very satisfactory results.
In this paper results are  presented from the radiative $F_2$ analysis         
($1.5 \le Q^2 \le 3.5$ GeV$^2$), from the  shifted vertex analysis  ($1.5 \le
Q^2 
\le $ 2.5 GeV$^2$),      
from a combination of the shifted vertex and the satellite bunch
 analysis         
($3.5 \le Q^2\le 6.5$ GeV$^2$),    
from the open triangle analysis ($Q^2$= 8.5 GeV$^2$)        
and from the nominal high statistics sample when the
scattered electron is detected in the BEMC   
($ 12 \le Q^2 \le 120~$GeV$^2$) or in the LAr calorimeter  
($ 120 < Q^2 \le 5000~$GeV$^2$).

Compared with the analysis of the 1993 data, 
many  uncertainties have been reduced.
 The systematic errors are due   
to the following sources:        
        
\begin{itemize}         
\item{The uncertainty in the electron energy scale which is 1\%        
in the BEMC, and  3\% in the LAr calorimeter. Since the $y_e$ 
resolution varies
as $1/y$ with the energy resolution even a 1\% error on $\delta E/E$ can
lead to 10\% errors on $F_2$ at low $y$ in the E method.
}     
\item
{The uncertainty in the hadronic energy scale: the detailed study of 
$y_h/y_e$ and of $p_{T,h}/p_{T,e}$ ($p_T$ is the momentum transverse to 
the beam axis) allowed the assignment of a 4\% error on the hadronic 
energy deposited in the LAr calorimeter, a 15\% error on the same quantity 
in the BEMC, and a 3\% error on the $y_h$ fraction carried by the tracks.
These errors take into account the intrinsic energy scale uncertainty
of each detector and the uncertainty of the sharing of the
total hadronic final state energy between these three subdetectors.
These numbers also include   
uncertainties due to the  treatment of the electronic noise 
in the LAr calorimeter      
 and the BEMC.}    
\item{An uncertainty of up to $1$~mrad for the electron polar angle which 
leads to an error on $F_2$ of 8\% at low $Q^2$.}         
\item{Apart from the electron identification,  
all efficiencies were determined from the data and compared with   
Monte Carlo simulation. The
agreement between the experimental and the simulated values for the   
individual efficiencies          
was found to be better than      
2\%. An overall error of 2\% was assigned due to    
the imperfect description of the various efficiencies.       
A larger error was added         
to account for the variation of the vertex reconstruction  
efficiency  at large $x$ (up to 8\%) 
where jets get closer to the beam pipe in the forward direction,          
and at small $x$ or large $\theta$  (up to 4\%) where H1     
had no further tracking device besides the BPC to monitor the         
vertex efficiency.}    
\item{
Uncertainties in the hadronic corrections, the      
cross section extrapolation      
towards $Q^2 = 0$ and higher order corrections, which give
an error of up to 2\% in the radiative correction.      
The accuracy was        
cross checked by comparing the HECTOR calculation  with      
the HERACLES Monte Carlo 
simulation results. The  agreement to the few percent        
level between  the structure function  results
obtained with the E and the $\Sigma$ methods is an additional cross check     
for the control of the radiative corrections.}
\item{The structure function dependence of the acceptance    
which was kept below
1\% by performing a two step iterative  analysis.  
The uncertainty in the simulation of the hadronic final state
reflects most prominently  in the efficiency for the
requirement of an interaction vertex from tracks.  
  A comparison of the different  models (sect 4.3) for  the hadronic     
final state  was used to assign an additional          
$3$\% systematic error entering in all analyses at low $x$      
through the vertex efficiency.} 
\item{Based on the control data sample of electron tagged %, genuine 
$\gamma p$  
events the uncertainty due to photoproduction background could be    
estimated to be smaller than 30\% of the correction applied.          
This is equivalent  to  at most a 5\% systematic error in       
 the highest $y$ bins at lower $Q^2$ only.}   
\item{Statistical errors in the Monte Carlo acceptance and   
efficiency calculations were computed and added quadratically         
to the  systematic error.}
\item{For the  analysis of radiative events an additional
1.5\% uncertainty on the photon energy measurement
in the photon tagger was considered and a 2\% systematic error was added
 due to the uncertainty of
the photon tagger geometrical acceptance.
An uncertainty on the trigger efficiency of 6\% to 9\% was
included for the lowest $x$ points.}
%; c) the radiative corrections
%have been estimated  using the LESKO program~\cite{LESKO} which includes
%higher orders QED radiation. The resulting systematic error is 
%varying between 7\% at the lowest $x,Q^2$ and 1\%.}
\end{itemize}

\begin{figure}[tp]    
\begin{center}          
\epsfig{file=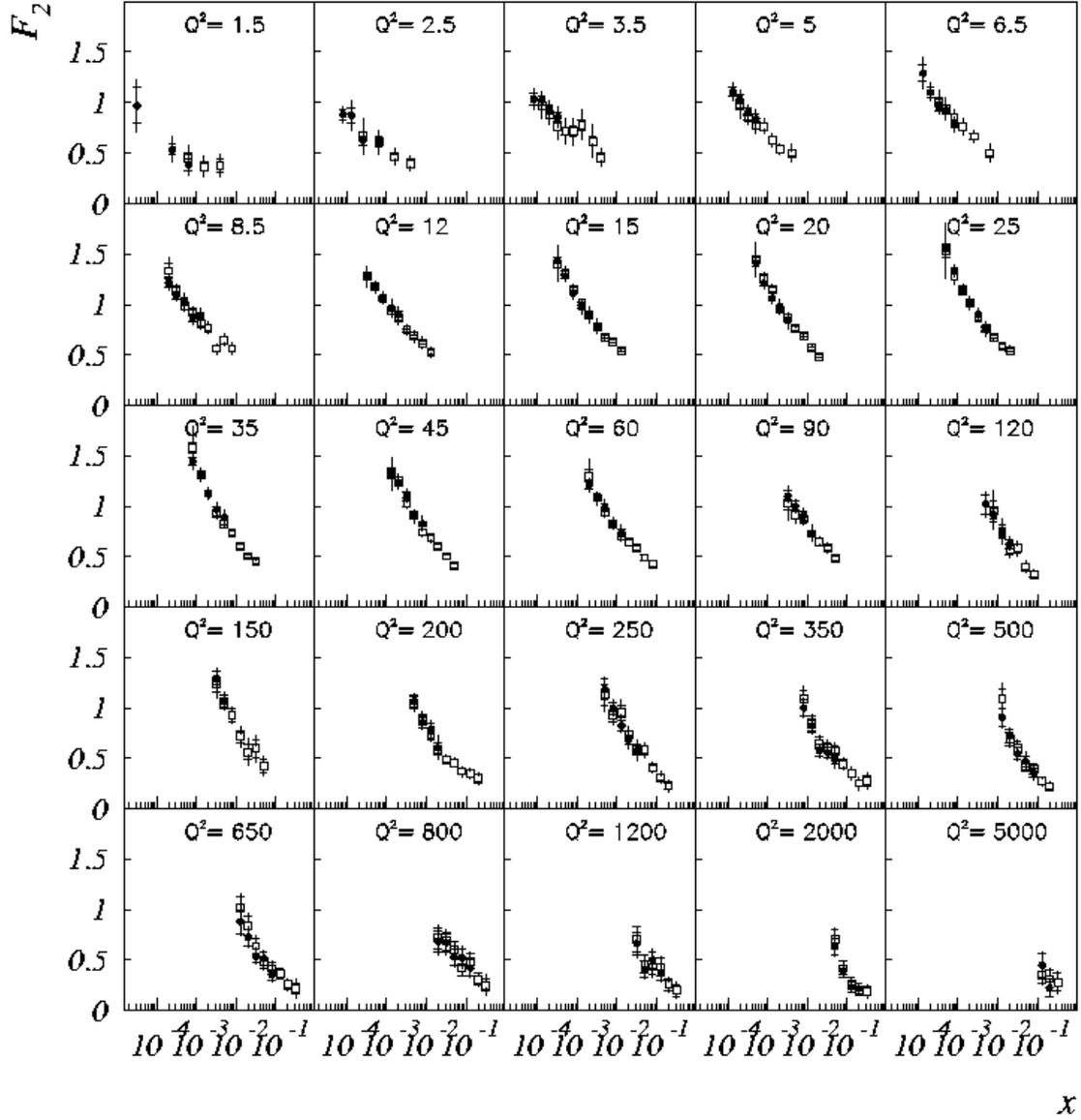,width=16cm,
bbllx=35pt,bblly=115pt,bburx=550pt,bbury=690pt}
\end{center}   
\caption[]{\label{f2xes}   
\sl  Measurement of the structure function with the electron (closed
circles) and the $\Sigma$ method (open squares). The inner error bar
is the statistical error. The full error bar represents the
statistical and systematic errors added in quadrature disregarding the
error from the luminosity measurement.}
\end{figure} 
  
\begin{figure}[htbp]    
\begin{center}          
\epsfig{file=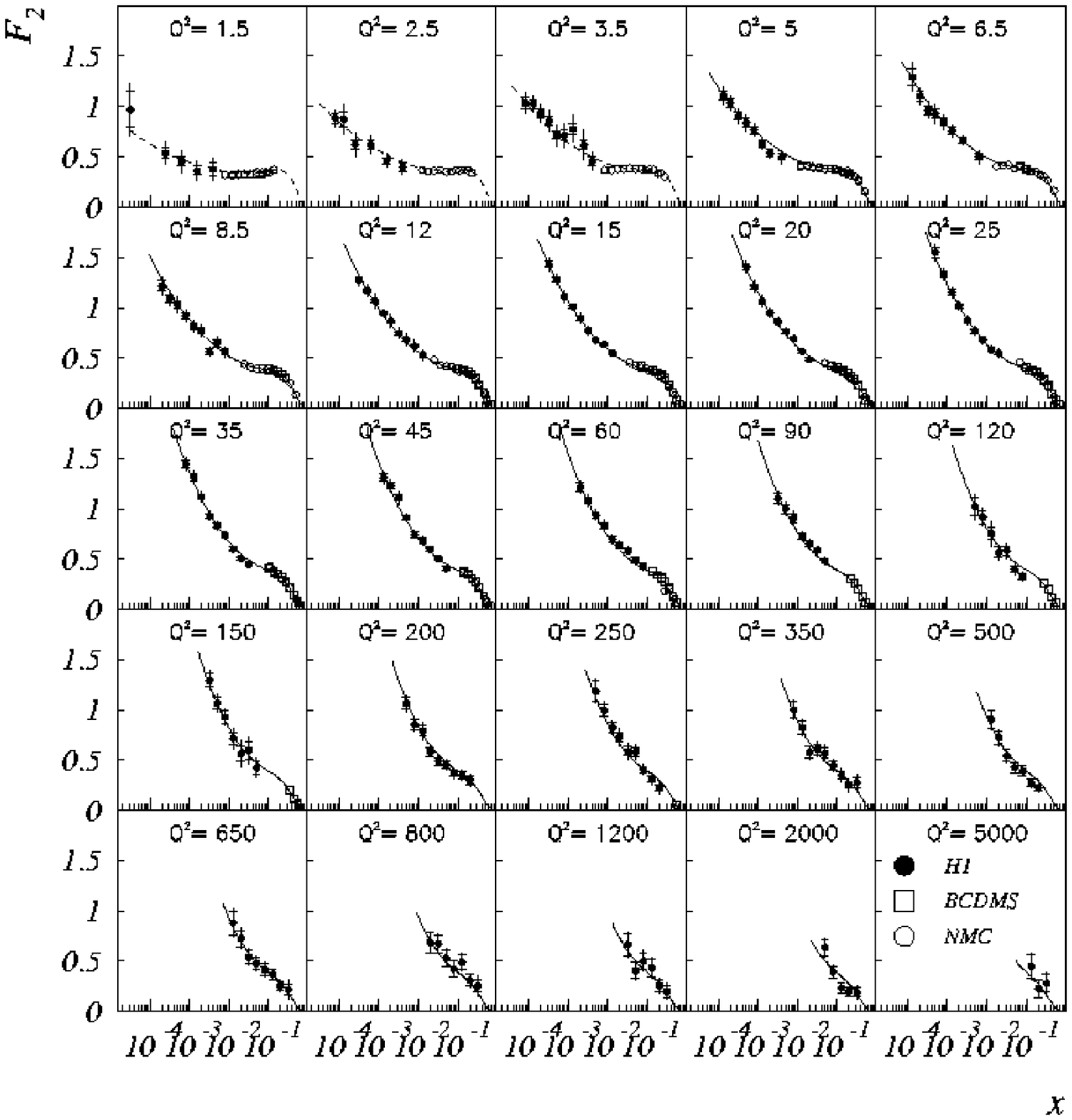,width=16cm,
  bbllx=15pt,bblly=140pt,bburx=550pt,bbury=690pt}
\end{center}   
\caption[]{\label{f2xall}   
{\sl  Measurement of the structure function $F_2(x,Q^2)$ as a function
of $x$. The closed
circles are the results of this analysis, the open circles are results taken
from the recent publication of the NMC~\cite{nmc} and the open squares
results from BCDMS~\cite{bcdms}. 
 The inner error bar
is the statistical error. The full error bar represents the
statistical and systematic errors added in quadrature and disregarding the
luminosity error.
Additionally a data point has also been measured at $Q^2=$ 2 GeV$^2$ (see
Table~\ref{f2tabl}, not shown in the Figure).
 The curves represent the NLO QCD fit described in section 7.4,
which includes the data for $Q^2 \geq$ 5 GeV$^2$.
 The extension of the curves below 5 GeV$^2$ represents only the
 backward evolution of the fit.}}
\end{figure}

Some of the systematic uncertainties affect differently the $F_2$  
measurement in the different methods. 
The systematic errors are given in Tables 7 and 8 point by point. However,
  some are strongly correlated over a large kinematic range.
  These correlations have been considered in the fits reported below.
  The matrix with the many different error contributions is available
  upon request to the H1 collaboration.
In Fig.~\ref{f2xes}    
the comparison of the measurements using  the  E and using the        
$\Sigma$ method is shown. Both are in good
agreement for all $Q^2$ values. %, indicating a correct estimation of the      
%systematic errors. 
%With the 1994 data, we can observe        
%that the strong rise  at low $x$      
%remains down to the lowest $x$ and $Q^2$ values reached in this analysis.
Some possible  discrepancies between both methods, e.g. at $Q^2 = 3.5$
  GeV$^2$, were investigated carefully and taken into account when
  evaluating the
  systematic error of the final structure function values if they could not
  be resolved. Small deviations are possible though, due to the different
  population of the $x,Q^2$ plane between  the methods of calculating the  
  kinematics.
 The two measurements are combined using the  E method for $y>0.15$     
and the $\Sigma$ method for $y<0.15$. The result  is shown in
Fig.~\ref{f2xall} and given in Tables 7 and 8.
The  measurements obtained from the low $Q^2$ nominal vertex sample
have a typical systematic error      
of  5\%. The large statistics     
allow  the measurement of  $F_2$  to reach      
  $Q^2$ values of  5000        
  GeV$^2$,     
and to achieve a few 
percent statistical precision at $Q^2$ below 100 GeV$^2$.     
%of the order of 2\%.
 The results are in good         
agreement with the previous H1 publication~\cite{H194}. 
In particular, the distinct rise of $F_2$ towards low $x$, observed    
with the 1992 and 1993~\cite{H193,H194} data, is confirmed with higher 
precision and extends now to significantly lower $Q^2$ values.
%extending even to very low $Q^2$ values. 
%In the $x,Q^2$ regions where both results exist, the 94 
%measurement is on average          
%about 5\% lower than the 93 one. Note that the      
%luminosity uncertainty for the 1993 data was 4.5\%,
% while now this has been    
%improved to the level of 1.5\%.  

\section{Discussion of the Results}

\subsection{Phenomenological Fits to the Data}

The $x$ and $Q^2$ behaviour of $F_2$ can be described by a
phenomenological ansatz of the type
\begin{equation}
F_2(x,Q^2) = [a \cdot x^b + c \cdot x^d \cdot ( 1 + e \cdot \sqrt x)
   \cdot (\log Q^2 + f \log ^2Q^2+h/Q^2)] \cdot (1-x)^g ,
\end{equation}
where $Q^2$ is given  in GeV$^2$.
This  functional form was introduced in detail 
previously \cite{H194}.
An extra $1/Q^2$ term has been added in order to get a good
description at  $Q^2$ below 5 GeV$^2$.
Note that this term is not a measure of higher twist contributions.
{\small
\begin{table} [htb!] \centering
\begin{tabular}{|c|c|c|c|c|c|c|c|}
\hline
       a & b & c & d & e & f & g  & h\\
\hline
   3.10 & 0.76 & 0.124 & $-$0.188 & $-$2.91 & $-$0.043 & 3.69 &1.40  \\
% II & 2.66 & 0.69 & 0.09 & $-$0.22 & $-$2.88 &  --   & 3.55  \\
\hline
\end{tabular}
\caption  {\label{tabpar}
 \sl Parameters of a phenomenological fit to the proton structure
function results from this experiment combined with $F_2$ 
measurements from the
NMC and the BCDMS experiments. The parametrization is valid for
1.5 GeV$^2 < Q^2 <$ 5000 GeV$^2$, 3$\cdot$10$^{-5} < x <$ 1 and
                                          $Q^2 < x \cdot$10$^5$ GeV$^2$.
The parameter $h$ is given in GeV$^2$.}
\end{table}}
For the fit,  results from  H1, NMC~\cite{nmc} and BCDMS~\cite{bcdms} are used
and statistical and systematic errors were added in quadrature.
The parameter values  quoted in Table~\ref{tabpar} are 
close to those obtained with  1993 data. 
%In the fit the statistical and
%systematic errors of all quoted $F_2$ values were added in quadrature.
%and the relative normalizations were not varied.
The fit provides   a good description of all data from the experiments
%considered here
 with a $\chi^2/dof$ of 1.65 
using full  errors. For the H1 data alone
the parametrization gives a $\chi^2/dof$ of 1.00.
%The parameters $g$ and $e$ come out as expected to
%be close to $\pm 3$. 
%The second term dominates at low $x$.
%
%The value of the power $d$ is found to be
%correlated with the presence of the $\log ^2Q^2$ term. If the latter is
%neglected the fit suggests an $x$ dependence at low $x$ which is
%somewhat steeper, namely $d=-0.23$ instead of $d=-0.19$.

\begin{figure}[htbp]                                                           
\begin{center}                                                                 
%\begin{picture}(0,0) \put(0,0){{\LARGE \bf  H1 1994}} \end{picture}
\epsfig{file=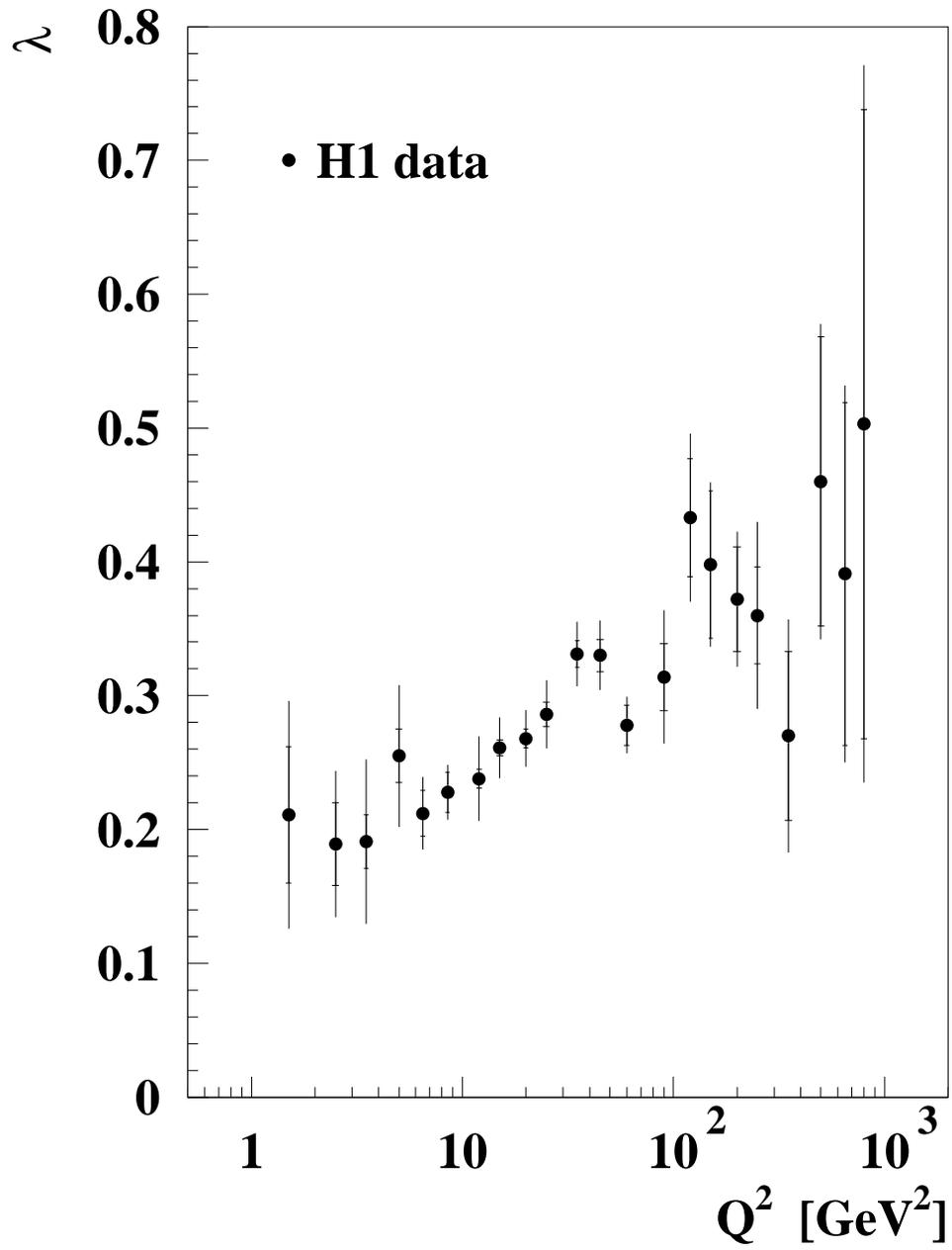,width=16cm,
  bbllx=30pt,bblly=50pt,bburx=520pt,bbury=585pt} 
\end{center}                                                                   
\caption[]{\label{expo}
\sl Variation  of the exponent $\lambda$   from fits
of the form $F_2 \sim x^{-\lambda}$ at fixed $Q^2$
values and $x<$ 0.1.} 
%The curves are fits of the type $a+b\log Q^2$.}
\end{figure}

In perturbative QCD the rate of growth of $F_2$ towards low $x$ is expected to increase  with increasing $Q^2$~\cite{Alvaro}. The wide range of
momentum transfer covered in this experiment enables a study of the
$Q^2$ dependence of the power $\lambda$ characterizing the  rise of $F_2$
$\propto x^{-\lambda}$ at low $x$. For each $Q^2$ bin and $x< 0.1$
the exponent
 $\lambda$ was determined taking into
 account the point-to-point systematic error correlations. The result
 is given in Table \ref{expotab} and displayed in Fig.~\ref{expo}. A
 rise of $\lambda$ with $\log Q^2$ is observed 
in the covered range from about 0.2 to 0.4 between 1.5 and 800 GeV$^2$. 

The structure function $F_2$ is
related to  the total cross-section of the virtual photon-proton
interaction, $\sigma_{tot}(\gamma^* p), $ via
\begin{equation}
 \sigma_{tot}(\gamma^* p) \simeq \frac{4~\pi^2 \alpha}{Q^2} F_2(W,Q^2).
\end{equation}
At low $x$,  $W$ is equal to  $\sqrt{Q^2/x}$.
 The $\lambda$ parameter thus  also determines
 the dependence of $F_2$ on  
%The observed rise of $F_2(x,Q^2)$ towards low $x$  means an increase
%of $F_2(W,Q^2)$ with rising $W$,
 the invariant mass squared  $W^2$ of the virtual
photon-proton ($\gamma^* p$) system. 
%due 
%to the relation
%\begin{equation}
% W^2 = Q^2  \cdot (1/x-1)+M^2_p \label{eqnw2}
%\end{equation}
%since $W^2$  for low $x$ %becomes simply $W^2 = 
%is equal to $Q^2/x$. 
%A linear extrapolation of the
%$Q^2$ dependence of  $\lambda$
%$\delta$ shows that $\delta$ is about 0.08
% for $Q^2 \simeq 0.3$ GeV$^2$ which is the  value expected for the
% $Q^2 \rightarrow 0$ limit of the behaviour of $F_2(W,Q^2)$ due to the
% relation of the structure function to the virtual photon-proton scattering
% cross section
%implies that  $\lambda=0.08$ for  $Q^2 $ about 0.2 GeV$^2$. 
For  hadronic  and real photoproduction total cross sections 
the value of  $\lambda$ has been measured to be around 0.08~\cite{dola2},
 which is
interpreted as the intercept of the so called soft pomeron.
For virtual  photon-proton interactions $\lambda$ is 
found to be substantially larger, and increases with $Q^2$. 
Future analyses of HERA data, which will lead to                 
 $F_2$ measurements at $Q^2$ below 1 GeV$^2$, should allow 
 the transition between 
deep-inelastic scattering  and real photoproduction to be studied.

%according to
%
%At fixed $Q^2$ the dependence of $F_2$ on $W^2$ was parametrized as
%$\sim W^{2 \delta}$
%
%              and fits were performed in the region of $x<0.1$ where
%$F_2$ rises taking into account the point-to-point systematic
%error correlations. In the covered $Q^2$ range starting at 1.5 $GeV^2$
%an approximately logarithmic dependence of the exponent $\delta$
%on $Q^2$ is observed as can be seen in Fig.~\ref{expo}
%and table \ref{expotab}.
%This parameter describes both the $W^2$ and the $x$ dependence
%since    eq.\ref{eqnw2} becomes simply
%  $W^2=Q^2/x$ for small $x$.
%Because of the relation of the structure function $F_2$ to the
%total $\gamma^* p$ cross section
%\begin{equation}
%\sigma_{tot}(\gamma^* p) \simeq \frac{4~\pi^2 \alpha}{Q^2} F_2(W,Q^2).
%\end{equation}
%the obtained result means that the rise of the cross section with
%$W$ becomes less steep with decreasing $Q^2$. The exponent
%$\delta$ approaches a value of 0.08 for $Q^2 = 0.3$~GeV$^2$ if
%the linear dependence is assumed to hold down to lower $Q^2$ values.

\begin{table}
\begin{center}
\begin{tabular}{|c|c|c|c||c|c|c|c|}
\hline
$Q^2/$GeV$^2$&$\lambda$&$\delta\lambda_{stat}$&$\delta\lambda_{syst}$&
$Q^2/$GeV$^2$&$\lambda$&$\delta\lambda_{stat}$&$\delta\lambda_{syst}$
\\
\hline
   1.5& 0.211& 0.051& 0.068&  45& 0.330& 0.012& 0.023\\
   2.5& 0.189& 0.031& 0.045&  60& 0.278& 0.015& 0.015\\
   3.5& 0.191& 0.020& 0.058&  90& 0.314& 0.025& 0.043\\
   5.0& 0.255& 0.020& 0.049& 120& 0.433& 0.044& 0.045\\
   6.5& 0.212& 0.017& 0.021& 150& 0.398& 0.055& 0.027\\
   8.5& 0.228& 0.015& 0.014& 200& 0.372& 0.039& 0.032\\
    12& 0.238& 0.007& 0.031& 250& 0.360& 0.036& 0.060\\
    15& 0.261& 0.006& 0.022& 350& 0.270& 0.063& 0.060\\
    20& 0.268& 0.007& 0.020& 500& 0.460& 0.108& 0.047\\
    25& 0.286& 0.009& 0.024& 650& 0.391& 0.128& 0.059\\
    35& 0.331& 0.010& 0.022& 800& 0.503& 0.235& 0.129\\
\hline
\end{tabular}
\caption{\sl  The values of the exponent $\lambda$ as a function
of $Q^2$.
}
\label{expotab}
\end{center}
\end{table}

%\section{Phenomenology}
\subsection{Comparison with Models at Low {\boldmath $Q^2$}} 
                                                   
\begin{figure}[htbp]    
\begin{center}          
\epsfig{file=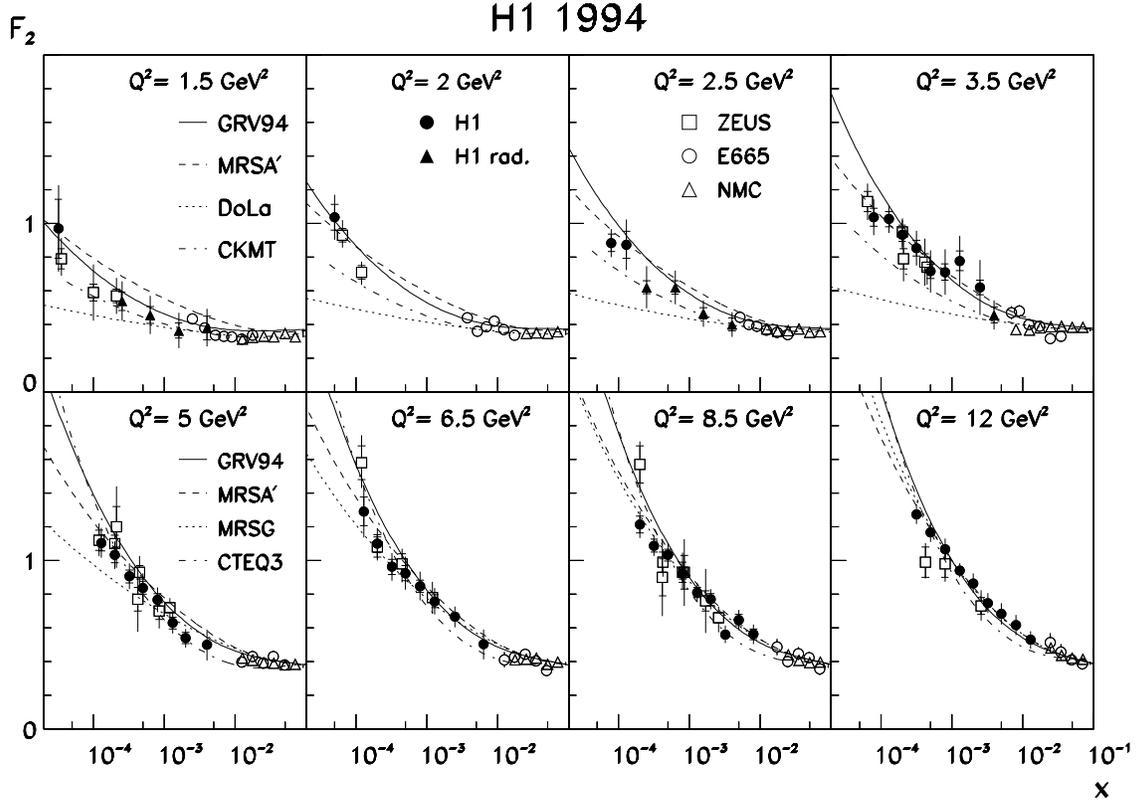,width=14cm,
 bbllx=60pt,bblly=200pt,bburx=530pt,bbury=600pt}
\end{center}   
\caption[]{\label{lowq2}   
\sl 
Measurement of the proton structure function $F_2(x,Q^2)$
in the low $Q^2$ region by H1 (closed circles: non-radiative events; 
closed triangles:  radiative events), together with results from the
ZEUS~\cite{zeuslowq2} (open squares),
E665~\cite{E665}
(open circles) and NMC~\cite{nmc} (open triangles) experiments.
The $Q^2$ values of the 
ZEUS data shown for the bins $Q^2= $3.5, 5 and 6.5 GeV$^2$ are 
measurements at 3.0, 4.5 and 6 $GeV$ respectively. 
Different parametrizations for $F_2$ are compared to the       
data. The DOLA and CKMT curves are only shown for the upper row of $Q^2$       
bins; CTEQ3M and MRSG are shown for the lower row; GRV 
and MRSA$^{\prime}$ are shown for the
full $Q^2$ range.}
\end{figure}

 Figures \ref{f2xall} and \ref{lowq2} clearly 
demonstrate the rise of $F_2$ with
decreasing $x$. In Fig.~\ref{lowq2} the  
data from  the eight lowest $Q^2$ bins 
  are shown\footnote{For the radiative data points (triangles in 
Fig.~\ref{lowq2})
 the $y$ variable cannot be calculated using $Q^2/sx$ with the nominal $s$
 since each bin has a different mean incoming electron energy. The average
 $y$ values are 0.143, 0.063, 0.026, 0.010 for $Q^2 = 1.5~$GeV$^2$  and
 0.199, 0.086, 0.036, 0.015
  for $Q^2=2.5~$GeV$^2$, starting at the smallest
 given $x$ value. For the single point at $Q^2=3.5~$GeV$^2$ $y=0.018$.}
and compared with
recent data and $F_2$ parametrizations. 
The rise of $F_2$ towards low $x$ is also present
in the low $Q^2$ region.
The measurement is in good agreement with the  data from  the ZEUS 
experiment~\cite{zeuslowq2} and matches well with the data 
 from fixed target experiments~\cite{E665,nmc} at higher $x$ values.

%In Fig.\ref{lowq2} t
The curves denoted as 
MRSA$^{\prime}$,
MRSG~\cite{mrsa}, CTEQ3M~\cite{cteq} and GRV~\cite{NEWGRV}
 are parametrizations based on the
conventional QCD evolution equations. 
   These calculations assume a certain shape
 of the $x$ behaviour   at a
 starting 
$Q^2_0$ value and use the DGLAP~\cite{DGLAP}
 equations to get predictions at different
 $Q^2$ values. The MRS and CTEQ distributions assume an 
  $ x^{-\lambda}$ behaviour for $ x \rightarrow 0$ at
 starting $Q^2_0$ of a few GeV$^2$.
 Their parameters  were determined
 using also the 1993 
HERA structure function data. 
%These started to be significant only at
%higher $Q^2$ values ($Q^2 \geq 7$ GeV$^2$).

The GRV calculation assumes  that all parton distributions at
very low  $Q^2_0= 0.34$ GeV$^2$
 have a valence like shape, i.e. vanish for $x\rightarrow 0$. 
Assuming that the DGLAP equations can be used
to evolve the parton distributions from
this low $Q^2_0$ scale to larger $Q^2$ values, they predicted that
   the structure
 function $F_2$ should rise towards low $x$ even for low values 
of  $Q^2\sim 1 $ GeV$^2$~\cite{NEWGRV}.
The determination of the shape parameters of the distributions at
the starting scale uses only data from fixed target experiments and not much
 freedom is left for further adjustments in the kinematic range of the
 HERA data. Small  variations are connected
 with  changes  still possible in  the starting $Q^2_0$ and the
value of the QCD parameter $\Lambda$.
Fig.~\ref{lowq2} shows that the GRV distributions describe the  data well,
indicating that in this kinematic regime  the sea quark 
distributions can be produced by QCD dynamics. 
%Despite the valence behaviour ansatz, which implies a decrease of
%the parton distributions for small $x$ at   $Q^2_0$,
%th%e parametrizations show a strong rise towards low $x$
%over the measured $Q^2$ range. This is due to  the long lever arm used in
%the DGLAP evolution from the starting scale $Q^2_0$
%to the measured
%$Q^2$ values. 
%The GRV~\cite{NEWGRV} distributions  contain the evolution
%of the charm quark distribution in NLO~\cite{RIEM}
% which affects
%mainly the lower $Q^2$
%region.

 Parametrizations motivated by Regge theory
 relate the structure function to
Reggeon exchange phenomena which successfully describe
 the slow rise of the total
cross section with the centre of mass system energy
in hadron-hadron and $\gamma p$ interactions.
Using the ``bare''
 instead of the ``effective'' pomeron
intercept,
 the CKMT~\cite{ckmt} parametrization
rises faster with $x$ compared to former DOLA~\cite{dola} calculations.
The CKMT curves 
were calculated using the parameters as given in~\cite{ckmt}, without 
QCD evolution in the whole range. 
%The effective  
%pomeron intercept $1+\delta$ amounts 
%$\delta = 0.20$ at $Q^2= 5$ GeV$^2$ 
% that $Q^2$ value.
%, shown below 5 GeV,
%hence they are compared to
%the data in the lowest $Q^2$ bins only.

The predictions for the 
 Regge inspired models DOLA and                                  
CKMT  lie below  the data for $Q^2 \ge 2 $ GeV at low $x$.
The latter  were already shown to be significantly below the H1 data of 
1993~\cite{deroeck}.
The
GRV and MRSA$^{\prime}$
 parametrizations give a good description of the data     
in the  range shown, with the possible exception of the first $Q^2$ 
bin for the latter.
The MRSG and CTEQ3 distributions, which are not available for 
the lowest $Q^2$ values,
describe the higher $Q^2$ data well.
% follow the general trend of the data.     

\subsection{Double Asymptotic Scaling}

The success of the GRV approach suggests that the observed rise of 
the structure function $F_2$ towards
low $x$ is generated by  QCD dynamics.
This was already observed in 1974~\cite{Alvaro} from a study of the 
 behaviour of $F_2$  in the limit of large $Q^2$ and 
low $x$. In this asymptotic region 
the QCD evolution determines the shape of $F_2$.
% equations become  simple and the shape of $F_2$ at small 
%$x$  is determined by radiative processes
%independent of the starting distribution.
Recently
Ball and Forte~\cite{ball}
% have revived this observation and 
developed  a convenient way to test 
the asymptotic
behaviour of  $F_2$ using  two variables
%$\sigma \sim \sqrt{\log (1/x) \log \log (Q^2)}$
%and   $\rho \sim \sqrt{\log (1/x) / \log \log (Q^2)}$. 
%They have also shown that large scaling violations can be a sign
%for the presence of BFKL effects in the structure function data.
%In the asymptotic region 
%$F_2$ is expected to scale in these two variables.
%This scaling behaviour
%of the HERA data was demonstrated in the region $x< 0.1, Q^2 > 5 $ GeV$^2$ 
%in~\cite{ball,qcdfit}.
%The theoretical expectations for this study were calculated  
%  at the one loop
%level~\cite{ball}. The two loop corrections can lead to substantial
%subasymptotic contributions at low $Q^2$, which can become visible
%at large $\rho$. Furthermore, at the one loop 
%level there is the usual  ambiguity on the choice 
%of $\Lambda$.
%Recently Ball and Forte have determined the  two loop corrections to 
%their calculations~\cite{bf2loop}, and the H1  data will be presented 
% using this  formalism. 
%The definitions of scaling variables 
%$\sigma$ and $\rho$  have now become: 
\begin{equation}
\sigma \equiv \sqrt{\log(x_0/x)\cdot \log(\alpha_s(Q_0)/\alpha_s(Q))}, \ \ \
 \rho \equiv \sqrt{\frac{\log(x_0/x)}{\log(\alpha_s(Q_0)/\alpha_s(Q))}}
\end{equation}
where $\alpha_s(Q)$ is evaluated at the 
two
loop level~\cite{bf2loop}. 

The parameters $x_0$ and $Q^2_0$ have to be determined 
experimentally. The parameter
$Q^2_0$ is optimized by minimizing the  $\chi^2$ 
of a linear fit of $\log(R_F'F_2)$ versus $\sigma$ (see below) using data with
$Q^2\ge$ 5 GeV$^2$. This leads to a value of $Q^2_0 = 2.5$ GeV$^2$.
%Note that this value is higher than that used for the one loop
%analysis~\cite{eisele}.
The same procedure was followed for $x_0$, which showed less sensitivity. 
The value  $x_0 = 0.1$, as suggested in~\cite{ball,qcdfit},
was found to be a good choice.
%In order to check the double asymptotic scaling
%H1 data are presented in the variables $\sigma$ and $\rho$. 
%taking
%the boundary conditions to be $x_0=0.1$ and 
%$Q^2_0=2.5~{\rm GeV}^2$,  and $\Lambda^{(4)}_{\rm LO}=289~{\rm MeV}$
To visualize the double scaling, it was
proposed to rescale $F_2$ 
%in order to 
%eliminate 
%part of the leading subasymptotic behaviour, with 
with factors $R_F'$ and $R_F$ 
defined as
\begin{equation}
R_F(\sigma,\rho) = 8.1\: \exp \left(-2\gamma\sigma+
\omega \frac{\sigma}{\rho}+\frac{1}{2}
\log(\gamma \sigma) +
\log (\frac{\rho}{\gamma})\right)/\xi_F
\end{equation}
with
\begin{equation}
\xi_F = 1+((\xi_1+\xi_2)*\alpha_s(Q)-\xi_1
*\alpha_s(Q_0))*(\rho/(2\pi*\gamma))
\end{equation}
and
\begin{equation}
R_F'(\sigma,\rho) = R_F\exp(2\gamma\sigma).
\end{equation}
Here
$\xi_1=(206n_f/27 + 6b_1/b_0)/b_0$,
$\xi_2=13$, $b_0=11-2n_f/3$,  $\omega = (11 + 2n_f/27)/b_0$ 
and $b_1=102-38n_f/3$. 
The number of flavours is $n_f$ 
 and $\gamma=\sqrt{(12/b_0)}$.
The function
$\log(R_F' F_2)$ is then predicted to rise linearly with $\sigma$.
%the same as in leading order:
%Here $\gamma \equiv 2\sqrt{3/b_0}$ with
%$b_0$ being the leading order coefficient of the $\beta$ function
%of the QCD renormalization group equation for four flavours, 
$R_F F_2$ is expected to be independent of 
  $\rho$ and $\sigma$.
%Deviations from this scaling behaviour 
%in the data could be an indication for e.g. BFKL like 
%contributions.
Note that these expectations are  valid only if 
the gluon distribution, which  drives $F_2$ at low
$x$ via the sea quarks, does not have a too singular behaviour for
%$Q^2$ values around and below $Q_0^2$.
$Q^2 = Q_0^2$.
%The slope of  $\log(R'_F*F_2)$ versus $\sigma$ is sensitive to 
%the value of $\alpha_s$. 

\begin{figure}[htbp]                                                           
\begin{center}                                                                 
\epsfig{file=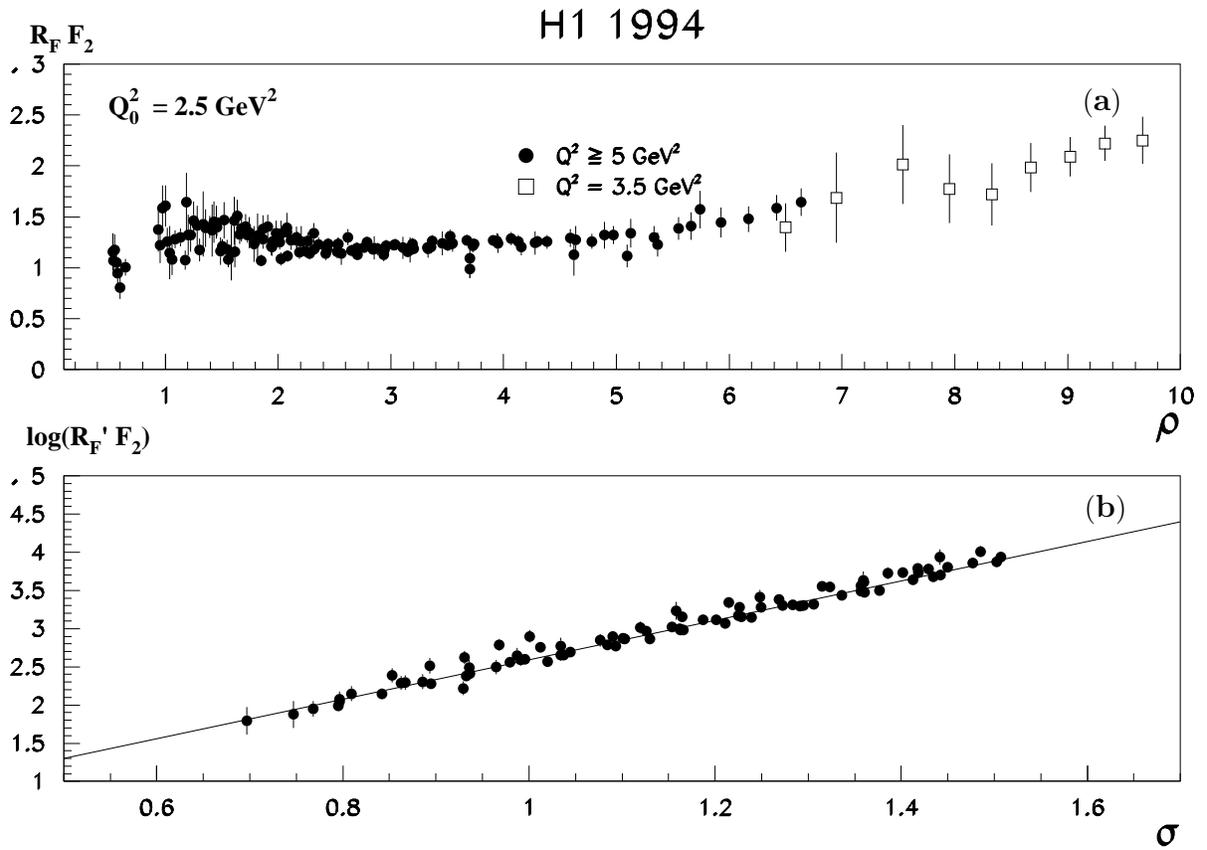,width=16cm,
  bbllx=35pt,bblly=235pt,bburx=540pt,bbury=595pt} 
\begin{picture}(0,0) \put(68,104){({\large \bf a})} \end{picture} 
\begin{picture}(0,0) \put(67,50){({\large \bf b})} \end{picture}
\end{center}                                                                   
\caption[]{\label{figball}                                                     
\sl 
The rescaled structure functions {\bf a)} $R_FF_2$ versus $\rho$ 
and {\bf b)}
$\log(R_F'F_2)$ versus $\sigma$ (see text).
Only data with $Q^2\ge$ 5  GeV$^2$ and $\rho>$ 2 are shown
in b).}
\end{figure}

Fig.~\ref{figball}a shows $R_FF_2$ versus $\rho$ for the 
data with $Q^2\ge 3.5$ GeV$^2$.
The value of 
$\Lambda$ for four flavours
was chosen to be $263$ MeV \cite{qcdbcd}.
%which corresponds to $\alpha_s(M_Z)= 0.117$.
 The continuity
of $\alpha_s(Q)$ at the bottom quark mass threshold is imposed using
the prescription in~\cite{marciano}.
Approximate scaling is observed for $Q^2 \ge 5$ GeV$^2$ and 
%Scaling roughly sets in for $Q^2 \ge 5$ GeV$^2$ and 
$\rho \ge 2$. 
At high $\rho$ the low $Q^2$ data tend to violate the scaling behaviour which
 is seen  clearly for the data at 3.5  GeV$^2$. 
%The scaling 
%violations  at these low $Q^2$ 
%values are expected to be due to either
%subasymptotic terms or non-perturbative effects. 
%rather than a signal for BFKL like contributions.
In Fig.~\ref{figball}b, the results are shown  for $\rho \ge 2$
and $Q^2\ge 5$ GeV$^2$ as a function of $\sigma$. The data exhibit the expected
 linear  rise of 
$\log(R_F' F_2)$  with $\sigma$.
A linear fit to the data gives a value for the slope of: 
 $2.50\pm 0.02 \pm 0.06$ 
($2.57\pm 0.05\pm0.06$) for $Q^2< 15$ GeV$^2$ ($Q^2>35$ GeV$^2$)
and 4 (5) flavours.
 The first error is the statistical error and
the 
second error is the  
systematic error taking into account the point-to-point correlations.
The value expected from QCD is 2.4 (2.5) for 4 (5) flavours. The results
are in agreement with these predictions. Compared to the 
result presented in ~\cite{qcdfit}, the extraction based on the 2-loop 
formalism used here is in better agreement with QCD expectation.
%It was determined in the following
%way: All systematic errors discussed in section~6 were used in turn to
%calculate a new set of $F_2$ values and a new fit was performed for each of these sets% of $F_2$ values. The differences of the new fit results and the 
%unshifted result were added quadratically to calculate the total systematic
%error.
Not included in this error is the influence of the uncertainty in the choice
of $\Lambda$. Varying $\Lambda$ by $\pm 65 $MeV changes the result
on the slope by $\mp 0.03$. 
 
%Using the newly available two loop calculations 
One can conclude 
that
the low $x$, low $Q^2$ measurements for  $Q^2 \ge 5$ GeV$^2$ show
scaling in $\rho$ and $\sigma$. 
%These data lie well  in the asymptotic region.  
Thus double asymptotic scaling is a dominant feature of $F_2$ in this region.
%, leaving little room for other contributions.

\begin{figure}[htbp]                                                           
\begin{center}                                                                
\epsfig{file=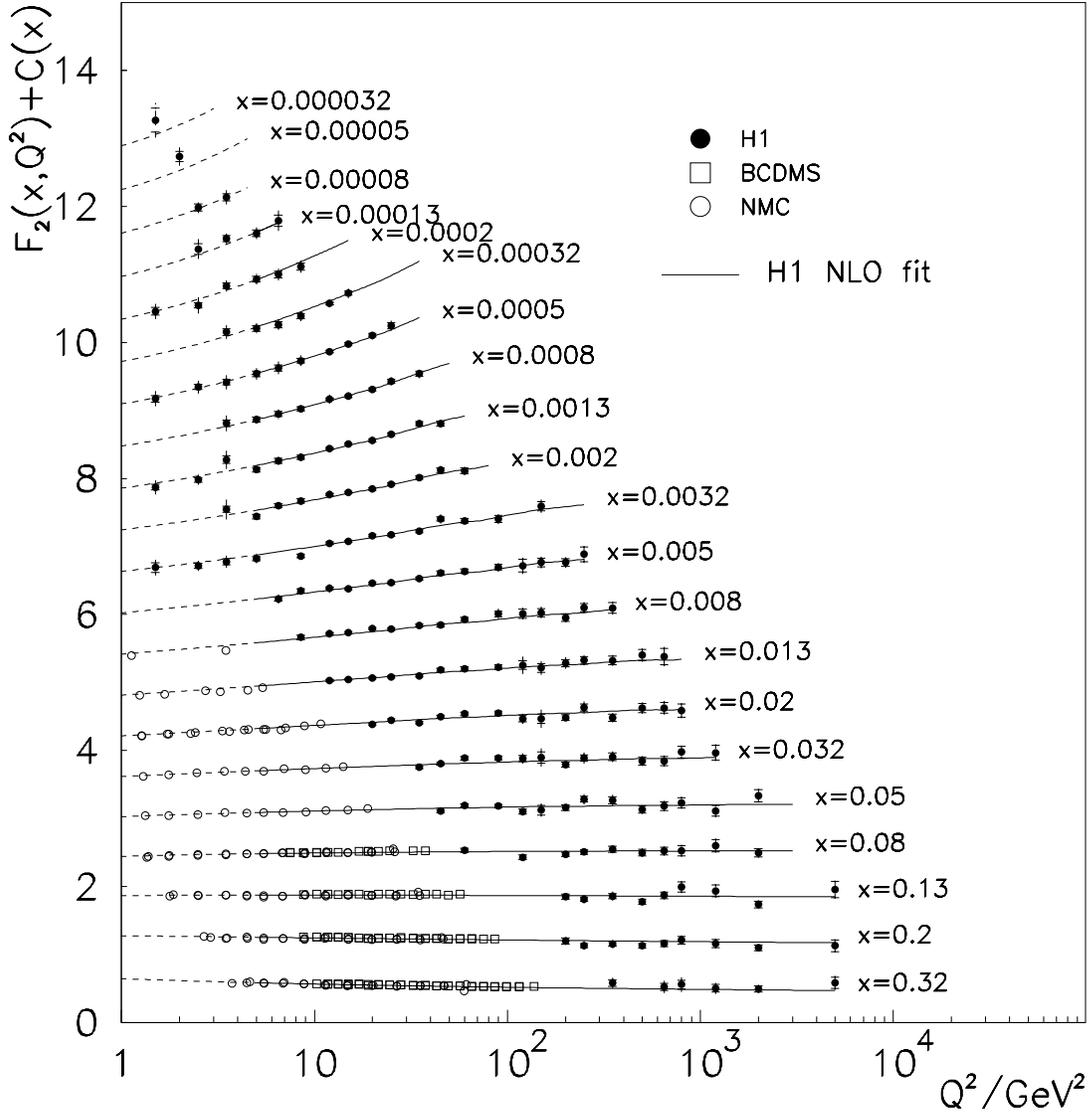,width=15cm,
  bbllx=15pt,bblly=140pt,bburx=570pt,bbury=690pt}
\end{center}                                                                  
\caption[]{\label{f2q2}                                                 
\sl $F_2(x,Q^2)$ measured by H1 together with BCDMS~\cite{bcdms} and
NMC~\cite{nmc} fixed target results.
The full line corresponds to the NLO
QCD fit, see sect.7.4, which includes the data for $Q^2 \geq$ 5 GeV$^2$.
 The extension of the curves below 5 GeV$^2$ represents only the
 backward evolution of the fit.
 The $F_2$ values are plotted in a linear scale adding 
 a  constant c($x$)= 
0.6(i-0.4) where $i$ is the $x$ bin number starting at $i=$1 from $x=$ 0.32.
The inner error bar is the statistical error, the outer corresponds to
the full error resulting from adding the statistical and systematic error
in quadrature.
Some H1 data points at lower $Q^2$ where shifted to 
nearby $x$ values
 for graphical representation of the data. 
 }
\end{figure}

\subsection{Extraction of the Gluon Density}
%The data presented here considerably extend the $(x,Q^2)$ range compared to
%the fixed target muon scattering experiments. 
%Thus it is of interest to extract with the improved accuracy of the
%$F_2$ measurement the gluon density, in particular at low $x$.
In a QCD analysis 
the evolution equations were solved numerically 
in the NLO order 
 approximation following the procedure described 
in~\cite{pascaud-lal}.
The splitting functions
% \cite{furman}
 and the strong coupling
constant $\alpha_s(Q^2)$
%~\cite{marciano}
 are defined in the $\overline{MS}$
factorization and renormalization
schemes. 
% \cite{furman2}.  
In the DGLAP evolution equations only three light quark 
flavours are taken into account. Heavy quark contributions are
dynamically generated 
using the photon-gluon fusion (PGF) prescription given in~\cite{grv1,grv2},
extended to   NLO according to \cite{RIEM}.
 The scale of the PGF process has been taken as
$\sqrt{Q^2+ 4 m_c^2}$ with a charm quark mass of $m_c=1.5$~GeV.  
An uncertainty of the charm quark mass of $0.5$~GeV was considered which
leads to a few percent variation of the gluon density. 
The  small contribution of beauty  quarks
has been  neglected.

 The gluon $g$, the valence  quark $u_v$
and $d_v$ and the non-strange sea $S$ ($S
\equiv \bar{u}+\bar{d}$) distributions are parametrized
at  $Q^2_0 = 5$ GeV$^2$  in the following way:
\begin{eqnarray}\label{input1}
xg(x)&=& A_gx^{B_g}(1-x)^{C_g},\nonumber \\
xu_v(x)&=& A_{u}x^{B_{u}}(1-x)^{C_{u}}(1+D_{u}x+E_{u}\sqrt{x}),\nonumber \\
xd_v(x)&=& A_{d} x^{B_{d} }(1-x)^{C_{d}}(1+D_{d}x+E_{d}\sqrt{x}),\nonumber \\
xS(x)&=& A_{S} x^{B_{S} }(1-x)^{C_{S}}(1+D_{S}x+E_{S}\sqrt{x}).
\end{eqnarray}
 The quark and antiquark components of the sea are assumed to be equal,
and $\bar{u}$ is set equal to $\bar{d}$. As determined in ~\cite{ccfr},
the strange quark density is taken to be $S/4$.
With these definitions  the proton structure function $F_2$,  for
$n_f=3$ and to leading order, is given as
\begin{equation}
 F_2(x,Q^2)= \frac{11}{18} xS +  \frac{4}{9} xu_v + \frac{1}{9} xd_v.
\end{equation}
%which in the kinematic range of the present H1 data reduces with good
%precision to the contribution from the sea distribution $S$.

%From these parton distributions, the singlet  
% $q_{SI}$ and two non-singlet densities 
% $q_{NS1}$ and  $q_{NS2}$
%%%%%%%%$q_{NS1}=u+\bar{u}-q_{SI}/3$, $q_{NS2}=u+\bar{u}-d-\bar{d}$ 
%are constructed according to the following relations:
% \begin{eqnarray}\label{input2}
%  q_{SI}(x)&=& u_v(x)+d_v(x)+2 S(x) (1+\frac{1}{4}),\nonumber \\
%  q_{NS1}(x)&=& \frac{2}{3}u_v(x)-\frac{1}{3}d_v(x)+(1-\frac{5}{6}) S(x)
%\nonumber \\
%  q_{NS2}(x)&=& u_v(x)-d_v(x).
% \end{eqnarray}

%The deuteron data better  constrain  the amount of valence quarks
%in the singlet distribution and the singlet part of the momentum sum rule. 
%Some parameters of eq.
%\ref{input1} are fixed using physical constraints: 

 The normalizations of
the valence quark densities are fixed using the counting rules 
 $\int_0^1u_v(x)dx=2$ and $\int_0^1d_v(x)dx=1$.
The normalization $A_{g}$ of the gluon density is obtained via
the momentum sum rule.
%%%%%~ $\int_0^1[xg(x)+xq_{SI}(x)]dx=1$
%%%%and only $A_g$ is determined by the fit. 
Since no isoscalar data are available yet in the small $x$ domain,
  $B_{u}=B_{d}$ is assumed. 
The parameters $B_S$ and $B_g$ which govern the 
small $x$ behaviour of $F_2$ and of the gluon are allowed to be
different.
For $\Lambda$  the value of 263 MeV is taken, as determined 
in~\cite{qcdbcd}. 

In order to constrain the valence quark densities at high $x$,
 proton and deuteron  results from the BCDMS and 
NMC    experiments are also used.
To avoid possible  higher twist effects,  data in the ranges  
$Q^2<5~{\rm GeV}^2$, and  $Q^2<15~{\rm GeV}^2$ for $x > 0.5$
are not included in the fit. The small contribution of 
large rapidity gap events in the HERA data
is considered to be part of the structure function, as there
is no evidence that the 
 QCD evolution of the diffractive part of
$F_2$ is significantly
different from that of the total inclusive $F_2$.

The parton densities
are derived from a fit of the evolution equations to the  data using the 
 program MINUIT. For the calculation of the $\chi^2$ which was minimized,  
the statistical errors were combined in quadrature with those systematic errors
which are uncorrelated. For BCDMS only statistical errors were 
included.
In addition a term was added to the 
$\chi^2$
to permit variation of  the relative normalization of the different data
sets.
%\footnote{For BCDMS, only statistical errors were 
%included. The $\chi^2$ obtained in this way was 931 for a total of 698 data
%points, which 
%can be decomposed into 187 units of $\chi^2$ (for 173 data points)
%for H1, 310 (192) for NMC, 430 (333) for BCDMS. 
% which only partially overlap in $x,Q^2$.
The following normalization errors were taken into account: 
H1 (nominal vertex sample): 1.5\%,  
H1 (shifted vertex sample): 3.9\%,
BCDMS: 3\%, and NMC: 2.5\%. 
The $\chi^2$ obtained in this procedure and the  $\chi^2$ computed when
considering the full error of each point are given in Table~\ref{tabchi}.
%A special treatment of the correlated systematic errors was
%done for the determination of the error on the gluon density.

The result of the fit is shown in Fig.~\ref{f2xall} 
versus $x$ and Fig.~\ref{f2q2} versus
$Q^2$. The fit gives a good description of all data used. Only
 small adjustments of the relative normalizations (given in Table
~\ref{tabchi}) are required demonstrating 
 remarkable agreement between these  different experiments. 
In Fig.~\ref{f2xall} the steep $x$ behaviour of $F_2$ is seen to 
be described very well by the fit. Note that the data for 
$Q^2 < 5 $~GeV$^2$, which were excluded from the fit, are still
 well reproduced by the fit evolved backwards in $Q^2$. 
However,   
%the $logQ^2$ variations in the very low $Q^2$ regions are big and the
a definite  test of perturbative QCD in this region requires 
more accurate data and a study of possible higher twist effects,
which is beyond the scope of this analysis. 
The $Q^2$ dependence at fixed $x$ is also described 
well over the nearly 4 orders of magnitude in $Q^2$ covered
by the H1 data, see Fig.~\ref{f2q2}.
 The parameters of the initial distributions are
listed in Table~\ref{tabqcd}. There are sizeable
correlations between these parameters which were not studied in detail as
 the basic aim of this analysis was  to extract 
the gluon density.

\begin{table}
\begin{center}
\begin{tabular}{|c|c|c|c|c|c|c|c|}
\hline
Experiment   &H1   & H1 &NMC-p&NMC-D&BCDMS-p&BCDMS-D&total\\
             & nvtx &  svtx & & & & & \\
\hline
 data points             & 157 & 16 &  96 &  96 & 174 & 159 & 698 \\
 $\chi^2$ (unco. err.)   & 174 & 13 & 157 & 153 & 222 & 208 & 931 \\
 $\chi^2$ (full error)   &  85 &  6 & 120 & 114 & 122 & 140 & 591 \\
normalization            & 1.00&1.04& 1.00&1.00 & 0.97& 0.97&      \\
\hline
\end{tabular}
\caption{ \sl For each experiment are given:
the number of data points used in the QCD fit, 
the $\chi^2$ obtained as described in the text
using only the uncorrelated errors, the $\chi^2$ computed from the same fit
using the full error on each point
and the normalization factors as determined from the fit.
The H1 nominal vertex  and shifted  vertex 
data samples are denoted as
nvtx and svtx respectively.}
\label{tabchi}
\end{center}
\end{table}

\begin{table}
\begin{center}
\begin{tabular}{|c|c|c|c|c|c|c|c|c|c|}
\hline
$A_g$&$B_g$&$C_g$&     &     &$A_u$ &$B_u$ &$C_u$ &$D_u$ & $E_u$\\
\hline
2.24& $-$0.20& 8.52&     &     &2.84  & 0.55 &4.19  & 4.42 & $-$1.40\\
\hline
$A_S$&$B_S$&$C_S$&$D_S$&$E_S$&$A_d$ &$B_d$ &$C_d$ &$D_d$ & $E_d$\\
\hline
0.27& $-$0.19& 1.66&0.16 &$-$1.00&1.05  & 0.55& 6.44 & $-$1.16& 3.87 \\
\hline
\end{tabular}
\caption{ \sl The values of the parameters at $Q^2=$5 GeV$^2$ of the
gluon, the sea quark  and valence
quark densities, as determined from the QCD fit.}
\label{tabqcd}
\end{center}
\end{table}

\begin{figure}[htbp]                                                          
\begin{center}                                                                
\begin{picture}(0,0) \put(75,156){{\LARGE \bf H1 1994}} \end{picture}
\epsfig{file=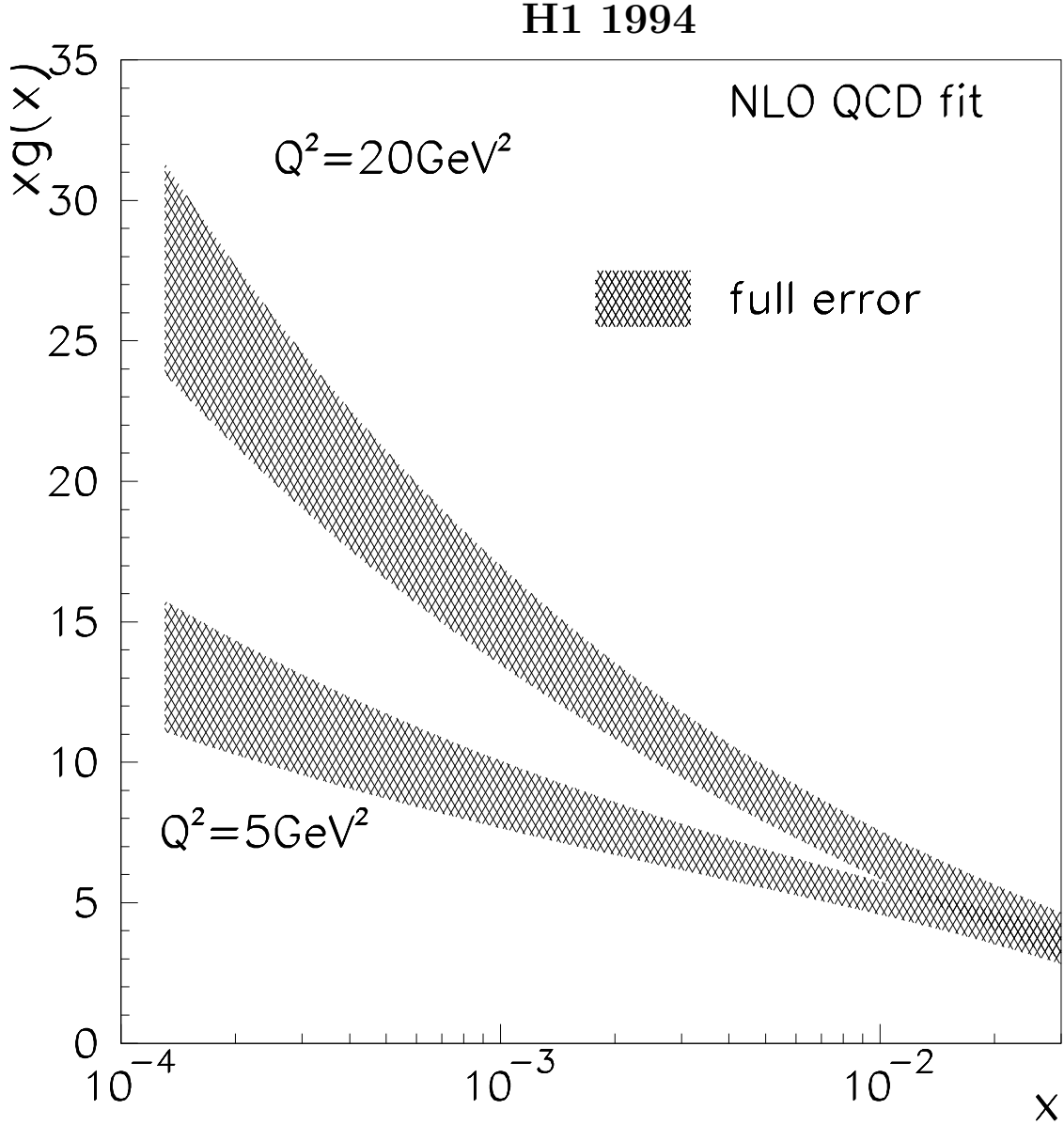,width=15cm,
  bbllx=15pt,bblly=145pt,bburx=550pt,bbury=695pt}
\end{center}                                                                  
\caption[]{\label{gluon}                                                 
\sl The gluon density $xg(x)$ at $Q^2 = $ 5 GeV$^2$ and 
$Q^2 = $ 20 GeV$^2$ extracted from a NLO  QCD
fit. The procedure to derive the error bands is explained in the text.}
\end{figure}

Fig.~\ref{gluon} shows the NLO gluon
 density $x g(x,Q^2)$ at $Q^2=5~{\rm GeV}^2$ and
$Q^2=20~{\rm GeV}^2$. Note that there are no $F_2$ 
measurements below $5\cdot 10^{-4}
$ at $Q^2= 20 $ GeV$^2$, but  in that region the gluon is constrained by the
data at lower $Q^2$ via the QCD evolution equations.
The  experimental error band was determined in  two steps: the initial
error was obtained directly 
from the fit when considering only 
the uncorrelated errors of the data points  which are dominated by
the statistical errors of the $F_2$ measurement. Then
the systematic errors introducing
point-to-point correlations, as 
for instance a possible shift in the scattered electron
energy, were treated  in a  procedure 
 described in~\cite{qcdfit}. Their effect
  was  added quadratically to the effect from the
uncorrelated ones, determining the full 
experimental error band of the measurement of $xg$.
% of this 
%measurement which is represented as the error band in the
%figure. 
A variation of 
$\Lambda$ by  65 MeV \cite{qcdbcd}
gives a  change of 9\%  on the gluon density 
at 20 GeV$^2$ which has not been added to the errors shown in Fig.~\ref{gluon}.
%sizeable and displayed separately:
%the external error bands are delimited by  the highest and lowest edges
%of the full error bands when the fit is performed using the 
%two extreme values $\Lambda=263+65$
%and $263-65$ MeV.  

% genuine measurement errors. At $Q^2=5~GeV^2$ an error
% $\delta \Lambda =\pm 65~MeV$ 
%\cite{qcdbcd} causes an uncertainty of 20\% for the gluon
% density at $x=10^{-4}$.
%The effect of a possible $\Lambda$ error
%on the gluon density is not small but kept separately from the
% genuine measurement errors. At $Q^2=5~GeV^2$ an error
% $\delta \Lambda =\pm 65~MeV$ 
%\cite{qcdbcd} causes an uncertainty of 20\% for the gluon
% density at $x=10^{-4}$.

The accuracy  of this determination of $xg$  
is better by  about a factor of two  than  the H1 result based on
the 1993 data~\cite{qcdfit}.
A  rise of the gluon density towards low $x$ is observed
which  is related  to the behaviour of $F_2 \propto x^{-\lambda}$.
Accordingly, the rise of $xg$ towards low $x$ increases with increasing $Q^2$.

\section{Summary}       
A measurement has been presented of       
 the proton structure function $F_2(x,Q^2)$ in      
deep-inelastic electron-proton scattering at HERA with data  
taken in the running period of 1994. The integrated          
luminosity is $2.7$~pb$^{-1}$ which represents  a tenfold    
increase in statistics compared to the $F_2$ analysis based on the 
1993 data of the H1 experiment.     
The structure function measurement includes  data from different      
detector components and running configurations. Low $Q^2$  values are 
reached using data with the $ep$ interaction vertex shifted  
from the nominal position and with radiative events.
 The data cover a kinematic range for $Q^2$ 
between $1.5$ and $5000$~GeV$^2$ and $x$ between $3.0 \cdot 10^{-5}$  
and $0.32$.    
        
The $F_2$ values presented       
are obtained using different methods to reconstruct the inclusive     
scattering kinematics. At high values of the scaling variable         
$y \geq 0.15$, due to its superior resolution,      
the  method  used is based on the scattered electron  
energy and angle.       
Lower $y$ values are covered with         
a method which combines electron and hadronic     
information to reduce radiative corrections and calibration  
errors. 
A smooth transition is observed from the fixed target high $x$ data 
to the HERA low $x$ data.        
        
The  rise  of the structure function with decreasing   
$x$ at fixed $Q^2$ is confirmed. The rate of growth increases with increasing $Q^2$
which has been one of the very first predictions of perturbative QCD.
%salient features of the very low $x$ behaviour of $F_2$ 
% and found to weaken logarithmically
% with decreasing $Q^2$, $F_2$ behaving like $x^{-0.16}$ at $Q^2 = 1$ GeV$^2$
% and like $x^{-0.27}$ at $Q^2 = 10$ $GeV^2$.
% Around $x \sim 10^{-3}$ the decrease of $x$ by an           
%order of magnitude amounts to a rise of $F_2$ 
%of about a factor two.  
%The rise is persistent at small $Q^2$ values, down to 2 GeV$^2$. 
%The structure function is proportional to $W^{2 \delta}$  
%where $W$  is the virtual ph%oton-proton centre--of--mass 
% energy with the exponent $\delta$
% also decreasing  nearly logarithmically with $Q^2$.
% a value around 0.1 at lowest $Q^2$ as expected from soft pomeron exchange.
% The steep rise of $F_2$ towards low $x$ is important in conforming the 
% observation of
 Approximate %double logarithmic 
scaling in  
double logarithmic  variables 
depending on $x$ and $Q^2$ is 
observed using a recent 2 loop  QCD calculation. %which has been a basic
% prediction of QCD and is studied here in NLO.     
The data are well described in the full $x$ and $Q^2$ range  by a NLO
 fit based on the conventional DGLAP evolution equations.   
The  fit results are used to measure the gluon distribution with improved
 precision down to $x =  10^{-4}$.
The gluon  density rises significantly for decreasing values of $x$.

\vspace*{1.cm}          
%\newpage      
%\noindent     
{\bf Acknowledgements}  
%=====================  
\normalsize    
        
\noindent      
 We are very grateful to the HERA machine group whose outstanding     
 efforts made this experiment possible. We acknowledge the support    
 of the DESY technical staff.    
 We appreciate the big effort of the engineers and           
 technicians who constructed and maintain the detector. We thank the         
 funding agencies for financial support of this experiment.  
     We wish          to thank the DESY    
 directorate for the support and hospitality extended to the          
 non-DESY members of the collaboration.  Finally, helpful discussions are acknowledged
with R. Ball, D. Bardin, J. Bl\"umlein, R.Peschanski, H. Spiesberger and A. Vogt.

\newpage
 
{\small
\begin{table}[t] \centering
\begin{minipage}[t]{7cm}
\begin{tabular}{|r|l|*{4}{c|}}
\hline
$Q^2$ & $x$  & $F_2$  & $\delta_{stat}$ & $\delta_{syst}$ & $R$  \\
\hline
    1.5 & .00003 & 0.969 & 0.176 & 0.187 & 0.71 \\
    1.5 & .00025 & 0.540 & 0.055 & 0.104 & 0.75 \\
    1.5 & .00063 & 0.458 & 0.050 & 0.101 & 0.74 \\
    1.5 & .00158 & 0.365 & 0.045 & 0.095 & 0.70 \\
    1.5 & .00398 & 0.381 & 0.070 & 0.087 & 0.63 \\
\hline
    2.0 & .00005 & 1.037 & 0.077 & 0.110 & 0.65 \\
\hline
    2.5 & .00008 & 0.885 & 0.052 & 0.065 & 0.80 \\
    2.5 & .00013 & 0.874 & 0.079 & 0.127 & 0.80 \\
    2.5 & .00025 & 0.622 & 0.037 & 0.119 & 0.80 \\
    2.5 & .00063 & 0.621 & 0.039 & 0.093 & 0.79 \\
    2.5 & .00158 & 0.466 & 0.033 & 0.072 & 0.75 \\
    2.5 & .00398 & 0.402 & 0.039 & 0.062 & 0.65 \\
\hline
    3.5 & .00008 & 1.036 & 0.053 & 0.092 & 0.64 \\
    3.5 & .00013 & 1.026 & 0.045 & 0.067 & 0.64 \\
    3.5 & .00020 & 0.934 & 0.041 & 0.075 & 0.64 \\
    3.5 & .00032 & 0.854 & 0.046 & 0.093 & 0.64 \\
    3.5 & .00050 & 0.716 & 0.041 & 0.119 & 0.64 \\
    3.5 & .00080 & 0.712 & 0.049 & 0.126 & 0.63 \\
    3.5 & .00130 & 0.778 & 0.058 & 0.137 & 0.61 \\
    3.5 & .00250 & 0.621 & 0.043 & 0.157 & 0.59 \\
    3.5 & .00398 & 0.458 & 0.046 & 0.075 & 0.54 \\
\hline
    5.0 & .00013 & 1.106 & 0.049 & 0.074 & 0.54 \\
    5.0 & .00020 & 1.033 & 0.044 & 0.069 & 0.54 \\
    5.0 & .00032 & 0.907 & 0.039 & 0.066 & 0.54 \\
    5.0 & .00050 & 0.839 & 0.039 & 0.076 & 0.53 \\
    5.0 & .00080 & 0.769 & 0.037 & 0.063 & 0.53 \\
    5.0 & .00130 & 0.630 & 0.034 & 0.050 & 0.51 \\
    5.0 & .00200 & 0.540 & 0.033 & 0.043 & 0.50 \\
    5.0 & .00400 & 0.500 & 0.029 & 0.086 & 0.46 \\
\hline
    6.5 & .00013 & 1.292 & 0.085 & 0.127 & 0.49 \\
    6.5 & .00020 & 1.101 & 0.052 & 0.072 & 0.48 \\
    6.5 & .00032 & 0.963 & 0.045 & 0.068 & 0.48 \\
    6.5 & .00050 & 0.926 & 0.044 & 0.088 & 0.48 \\
    6.5 & .00080 & 0.848 & 0.038 & 0.076 & 0.47 \\
    6.5 & .00130 & 0.759 & 0.039 & 0.068 & 0.46 \\
    6.5 & .00250 & 0.667 & 0.029 & 0.054 & 0.43 \\
    6.5 & .00630 & 0.504 & 0.029 & 0.084 & 0.37 \\
\hline
    8.5 & .00020 & 1.215 & 0.050 & 0.062 & 0.44 \\
    8.5 & .00032 & 1.089 & 0.038 & 0.048 & 0.44 \\
    8.5 & .00050 & 1.033 & 0.034 & 0.062 & 0.43 \\
    8.5 & .00080 & 0.923 & 0.031 & 0.038 & 0.43 \\
    8.5 & .00130 & 0.811 & 0.030 & 0.047 & 0.42 \\
    8.5 & .00200 & 0.770 & 0.034 & 0.049 & 0.40 \\
    8.5 & .00320 & 0.562 & 0.028 & 0.043 & 0.38 \\
    8.5 & .00500 & 0.648 & 0.033 & 0.051 & 0.36 \\
    8.5 & .00800 & 0.564 & 0.032 & 0.049 & 0.33 \\
\hline
   12. & .00032 & 1.276 & 0.020 & 0.055 & 0.39 \\
   12. & .00050 & 1.168 & 0.016 & 0.056 & 0.39 \\
   12. & .00080 & 1.067 & 0.015 & 0.061 & 0.38 \\
\hline
\end{tabular}\end{minipage}\hfill
\begin{minipage}[t]{7cm}
\begin{tabular}{|c|l|*{4}{c|}}
\hline
$Q^2$ & $x$  & $F_2$  & $\delta_{stat}$ & $\delta_{syst}$ & $R$  \\
\hline
   12. & .00130 & 0.942 & 0.015 & 0.039 & 0.37 \\
   12. & .00200 & 0.866 & 0.016 & 0.057 & 0.36 \\
   12. & .00320 & 0.749 & 0.016 & 0.055 & 0.34 \\
   12. & .00500 & 0.685 & 0.016 & 0.061 & 0.32 \\
   12. & .00800 & 0.618 & 0.016 & 0.057 & 0.30 \\
   12. & .01300 & 0.531 & 0.017 & 0.049 & 0.26 \\
\hline
   15. & .00032 & 1.426 & 0.030 & 0.064 & 0.37 \\
   15. & .00050 & 1.280 & 0.020 & 0.050 & 0.36 \\
   15. & .00080 & 1.110 & 0.018 & 0.057 & 0.35 \\
   15. & .00130 & 1.008 & 0.016 & 0.033 & 0.35 \\
   15. & .00200 & 0.895 & 0.015 & 0.046 & 0.34 \\
   15. & .00320 & 0.773 & 0.014 & 0.036 & 0.32 \\
   15. & .00500 & 0.677 & 0.014 & 0.035 & 0.30 \\
   15. & .00800 & 0.634 & 0.014 & 0.031 & 0.28 \\
   15. & .01300 & 0.547 & 0.013 & 0.027 & 0.24 \\
\hline
   20. & .0005 & 1.407 & 0.026 & 0.054 & 0.34 \\
   20. & .0008 & 1.210 & 0.022 & 0.050 & 0.33 \\
   20. & .0013 & 1.061 & 0.020 & 0.055 & 0.33 \\
   20. & .0020 & 0.945 & 0.018 & 0.042 & 0.32 \\
   20. & .0032 & 0.861 & 0.017 & 0.038 & 0.31 \\
   20. & .0050 & 0.761 & 0.017 & 0.028 & 0.30 \\
   20. & .0080 & 0.693 & 0.016 & 0.035 & 0.28 \\
   20. & .0130 & 0.567 & 0.015 & 0.024 & 0.26 \\
   20. & .0200 & 0.487 & 0.015 & 0.025 & 0.22 \\
\hline
   25. & .0005 & 1.546 & 0.047 & 0.058 & 0.40 \\
   25. & .0008 & 1.330 & 0.028 & 0.051 & 0.39 \\
   25. & .0013 & 1.151 & 0.024 & 0.047 & 0.38 \\
   25. & .0020 & 1.019 & 0.022 & 0.035 & 0.37 \\
   25. & .0032 & 0.872 & 0.020 & 0.034 & 0.35 \\
   25. & .0050 & 0.768 & 0.019 & 0.034 & 0.33 \\
   25. & .0080 & 0.683 & 0.018 & 0.031 & 0.30 \\
   25. & .0130 & 0.585 & 0.017 & 0.028 & 0.26 \\
   25. & .0200 & 0.548 & 0.017 & 0.037 & 0.22 \\
\hline
   35. & .0008 & 1.442 & 0.038 & 0.051 & 0.36 \\
   35. & .0013 & 1.308 & 0.032 & 0.052 & 0.35 \\
   35. & .0020 & 1.116 & 0.027 & 0.052 & 0.33 \\
   35. & .0032 & 0.928 & 0.024 & 0.038 & 0.32 \\
   35. & .0050 & 0.832 & 0.023 & 0.040 & 0.30 \\
   35. & .0080 & 0.739 & 0.022 & 0.035 & 0.27 \\
   35. & .0130 & 0.600 & 0.019 & 0.025 & 0.24 \\
   35. & .0200 & 0.508 & 0.019 & 0.019 & 0.20 \\
   35  & .0320 & 0.452 & 0.019 & 0.026 & 0.16 \\
\hline
   45. & .0013 & 1.305 & 0.038 & 0.048 & 0.32 \\
   45. & .0020 & 1.225 & 0.034 & 0.049 & 0.31 \\
   45. & .0032 & 1.105 & 0.032 & 0.058 & 0.30 \\
   45. & .0050 & 0.912 & 0.028 & 0.033 & 0.28 \\
   45. & .0080 & 0.743 & 0.025 & 0.029 & 0.26 \\
   45. & .0130 & 0.686 & 0.024 & 0.031 & 0.22 \\
\hline
\end{tabular}\end{minipage}
\caption {\label{f2tabl}
\sl  Proton structure function $F_2(x,Q^2)$ with statistical and
 systematic errors, part I.
The normalization uncertainty, not included in the systematic error,
 is 1.5\% for $Q^2 \ge$ 8.5 GeV$^2$ and 3.9\% for $Q^2 <$ 8.5 GeV$^2$. } 
% errors at lower $Q^2$.}
%                  All points have an additional scale uncertainty of
%1.5\% due to the luminosity determination. $Q^2$ is given in $GeV^2$.
%R was calculated according to QCD with a higher twist term added.}
\end{table}}

\newpage
{\small
\begin{table}[tp] \centering
\begin{minipage}[t]{7cm}
\begin{tabular}{|r|l|*{4}{c|}}
\hline
$Q^2$ & $x$  & $F_2$  & $\delta_{stat}$ & $\delta_{syst}$ & $R$  \\
\hline
   45. & .0200 & 0.599 & 0.022 & 0.027 & 0.19 \\
   45. & .0320 & 0.505 & 0.021 & 0.023 & 0.15 \\
   45. & .0500 & 0.411 & 0.022 & 0.028 & 0.12 \\
\hline
   60. & .0020 & 1.213 & 0.042 & 0.048 & 0.29 \\
   60. & .0032 & 1.079 & 0.037 & 0.045 & 0.28 \\
   60. & .0050 & 0.937 & 0.033 & 0.043 & 0.26 \\
   60. & .0080 & 0.830 & 0.031 & 0.046 & 0.24 \\
   60. & .0130 & 0.701 & 0.028 & 0.029 & 0.21 \\
   60. & .0200 & 0.639 & 0.027 & 0.025 & 0.18 \\
   60. & .0320 & 0.586 & 0.026 & 0.028 & 0.14 \\
   60. & .0500 & 0.492 & 0.025 & 0.023 & 0.11 \\
   60. & .0800 & 0.432 & 0.027 & 0.023 & 0.08 \\
\hline
   90. & .0032 & 1.103 & 0.052 & 0.048 & 0.26 \\
   90. & .0050 & 0.997 & 0.045 & 0.047 & 0.24 \\
   90. & .0080 & 0.908 & 0.041 & 0.056 & 0.22 \\
   90. & .0130 & 0.726 & 0.035 & 0.040 & 0.19 \\
   90. & .0200 & 0.650 & 0.033 & 0.031 & 0.17 \\
   90. & .0320 & 0.587 & 0.030 & 0.034 & 0.13 \\
   90. & .0500 & 0.481 & 0.027 & 0.019 & 0.10 \\
\hline
  120. & .0050 & 1.018 & 0.094 & 0.076 & 0.23 \\
  120. & .0080 & 0.914 & 0.068 & 0.056 & 0.21 \\
  120. & .0130 & 0.755 & 0.063 & 0.111 & 0.18 \\
  120. & .0200 & 0.570 & 0.049 & 0.057 & 0.16 \\
  120. & .0320 & 0.582 & 0.048 & 0.060 & 0.13 \\
  120. & .0500 & 0.402 & 0.035 & 0.045 & 0.10 \\
  120. & .0800 & 0.330 & 0.032 & 0.034 & 0.07 \\
\hline
  150. & .0032 & 1.292 & 0.069 & 0.067 & 0.23 \\
  150. & .0050 & 1.067 & 0.065 & 0.057 & 0.22 \\
  150. & .0080 & 0.928 & 0.061 & 0.060 & 0.20 \\
  150. & .0130 & 0.716 & 0.064 & 0.079 & 0.17 \\
  150. & .0200 & 0.566 & 0.069 & 0.114 & 0.15 \\
  150. & .0320 & 0.598 & 0.085 & 0.103 & 0.12 \\
  150. & .0500 & 0.424 & 0.071 & 0.065 & 0.09 \\
\hline
  200. & .005 & 1.065 & 0.059 & 0.053 & 0.21 \\
  200. & .008 & 0.853 & 0.051 & 0.038 & 0.19 \\
  200. & .013 & 0.787 & 0.052 & 0.071 & 0.17 \\
  200. & .020 & 0.585 & 0.041 & 0.023 & 0.14 \\
  200. & .032 & 0.490 & 0.038 & 0.026 & 0.11 \\
  200. & .050 & 0.460 & 0.039 & 0.029 & 0.09 \\
  200. & .080 & 0.372 & 0.032 & 0.039 & 0.06 \\
  200. & .130 & 0.350 & 0.037 & 0.032 & 0.04 \\
  200. & .200 & 0.301 & 0.045 & 0.036 & 0.03 \\
\hline
  250. & .005 & 1.185 & 0.106 & 0.060 & 0.20 \\
  250. & .008 & 1.000 & 0.062 & 0.054 & 0.18 \\
  250. & .013 & 0.826 & 0.055 & 0.047 & 0.16 \\
  250. & .020 & 0.730 & 0.051 & 0.072 & 0.14 \\
  250. & .032 & 0.590 & 0.044 & 0.067 & 0.11 \\
  250. & .050 & 0.584 & 0.043 & 0.060 & 0.09 \\
\hline
\end{tabular}\end{minipage}\hfill
\begin{minipage}[t]{7cm}
\begin{tabular}{|c|l|*{4}{c|}}
\hline
$Q^2$ & $x$  & $F_2$  & $\delta_{stat}$ & $\delta_{syst}$ & $R$  \\
\hline
  250. & .080 & 0.408 & 0.033 & 0.037 & 0.06 \\
  250. & .130 & 0.312 & 0.029 & 0.051 & 0.04 \\
  250. & .200 & 0.231 & 0.031 & 0.056 & 0.03 \\
\hline
  350. & .008 & 0.997 & 0.082 & 0.049 & 0.17 \\
  350. & .013 & 0.825 & 0.066 & 0.043 & 0.15 \\
  350. & .020 & 0.581 & 0.052 & 0.042 & 0.13 \\
  350. & .032 & 0.608 & 0.054 & 0.056 & 0.10 \\
  350. & .050 & 0.570 & 0.052 & 0.061 & 0.08 \\
  350. & .080 & 0.447 & 0.043 & 0.038 & 0.06 \\
  350. & .130 & 0.356 & 0.036 & 0.057 & 0.04 \\
  350. & .200 & 0.256 & 0.036 & 0.055 & 0.03 \\
  350. & .320 & 0.280 & 0.051 & 0.061 & 0.02 \\
\hline
  500. & .013 & 0.904 & 0.083 & 0.050 & 0.14 \\
  500. & .020 & 0.725 & 0.065 & 0.046 & 0.12 \\
  500. & .032 & 0.546 & 0.059 & 0.034 & 0.10 \\
  500. & .050 & 0.433 & 0.051 & 0.035 & 0.08 \\
  500. & .080 & 0.397 & 0.047 & 0.032 & 0.06 \\
  500. & .130 & 0.276 & 0.036 & 0.030 & 0.04 \\
  500. & .200 & 0.228 & 0.035 & 0.027 & 0.03 \\
\hline
  650. & .013 & 0.881 & 0.120 & 0.076 & 0.14 \\
  650. & .020 & 0.727 & 0.081 & 0.061 & 0.12 \\
  650. & .032 & 0.545 & 0.068 & 0.047 & 0.09 \\
  650. & .050 & 0.483 & 0.062 & 0.045 & 0.07 \\
  650. & .080 & 0.422 & 0.059 & 0.031 & 0.05 \\
  650. & .130 & 0.369 & 0.050 & 0.030 & 0.04 \\
  650. & .200 & 0.262 & 0.044 & 0.042 & 0.03 \\
  650. & .320 & 0.222 & 0.055 & 0.074 & 0.02 \\
\hline
  800. & .020 & 0.686 & 0.098 & 0.083 & 0.11 \\
  800. & .032 & 0.676 & 0.085 & 0.082 & 0.09 \\
  800. & .050 & 0.533 & 0.075 & 0.067 & 0.07 \\
  800. & .080 & 0.428 & 0.075 & 0.057 & 0.05 \\
  800. & .130 & 0.490 & 0.075 & 0.066 & 0.04 \\
  800. & .200 & 0.312 & 0.057 & 0.073 & 0.03 \\
  800. & .320 & 0.258 & 0.065 & 0.090 & 0.02 \\
\hline
 1200. & .032 & 0.668 & 0.109 & 0.091 & 0.09 \\
 1200. & .050 & 0.412 & 0.078 & 0.064 & 0.07 \\
 1200. & .080 & 0.502 & 0.089 & 0.069 & 0.05 \\
 1200. & .130 & 0.436 & 0.084 & 0.066 & 0.03 \\
 1200. & .200 & 0.260 & 0.057 & 0.048 & 0.02 \\
 1200. & .320 & 0.201 & 0.056 & 0.064 & 0.01 \\
\hline
 2000. & .05 & 0.634 & 0.087 & 0.046 & 0.06 \\
 2000. & .08 & 0.395 & 0.060 & 0.035 & 0.05 \\
 2000. & .13 & 0.237 & 0.048 & 0.026 & 0.03 \\
 2000. & .20 & 0.199 & 0.041 & 0.017 & 0.02 \\
 2000. & .32 & 0.193 & 0.043 & 0.045 & 0.01 \\
\hline
 5000. & .13 & 0.453 & 0.121 & 0.056 & 0.03 \\
 5000. & .20 & 0.229 & 0.087 & 0.030 & 0.02 \\
 5000. & .32 & 0.283 & 0.085 & 0.064 & 0.01 \\
\hline
\end{tabular}\end{minipage}
\caption {\label{f2tabh}  
\sl  Proton structure function $F_2(x,Q^2)$ with statistical and
 systematic errors, part II. 
The normalization uncertainty, not included in the systematic error,
 is 1.5\%.}
% at larger $Q^2$.}
%All points have an additional scale uncertainty of
%1.5\% due to the luminosity determination. $Q^2$ is given in $GeV^2$.
%R was calculated according to QCD with a higher twist term added.}
\end{table}}

\end{document}